\documentclass[journal]{IEEEtran}
\usepackage{graphicx}
\usepackage{color}
\usepackage{placeins}
\usepackage{float}
\usepackage{tabularx,colortbl}
\usepackage{graphics, subfigure, theorem, times, amsfonts, amsmath, amssymb, cite, verbatim}
\usepackage{multicol,multirow,booktabs}
\usepackage{tikz}
\usepackage{framed}
\usetikzlibrary{shapes,arrows}
\tikzstyle{phantom vertex} = [ ellipse, 
                               anchor = center, 
                               minimum height = 1*\unit, 
                               minimum width  = 1*\unit,
                               inner sep=0pt,
                               anchor=center]
\tikzstyle{red vertex}   = [black, fill = red!20,   phantom vertex, draw]
\tikzstyle{black vertex} = [black, fill = black!20, phantom vertex, draw]
\tikzstyle{blue vertex}  = [black, fill = blue!20,  phantom vertex, draw]
\tikzstyle{green vertex} = [black, fill = green!20,  phantom vertex, draw]
\tikzstyle{yellow vertex} = [black, fill = yellow!20,  phantom vertex, draw]
\tikzstyle{vertex}       = [draw, phantom vertex]

\tikzstyle{point} = [ellipse, inner sep=0pt, draw, fill=white, anchor = center,
                     minimum height = 0.05*\unit, minimum width  = 0.05*\unit]

\usepackage{pgfplots}
\usepackage{color}
\usepackage{hyperref}

\usepackage{needspace}

\newenvironment{myproof}[1][$\!\!$]{{\noindent\bf Proof #1: }}
                         {\hfill\QED\medskip}

\newcounter{excercise}
\newcounter{excercisepart}

\definecolor{pennblue}{cmyk}{1,0.65,0,0.30}
\definecolor{pennred}{cmyk}{0,1,0.65,0.34}
\definecolor{mygreen}{rgb}{0.10,0.50,0.10}

\def \reals    {{\mathbb R}}

\def\ccalC{{\ensuremath{\mathcal C}}}
\def\ccalD{{\ensuremath{\mathcal D}}}

\def\ccalH{{\ensuremath{\mathcal H}}}

\def\ccalL{{\ensuremath{\mathcal L}}}

\def\ccalN{{\ensuremath{\mathcal N}}}

\def\ccalP{{\ensuremath{\mathcal P}}}
\def\ccalQ{{\ensuremath{\mathcal Q}}}

\def\ccal0{{\ensuremath{\mathcal 0}}}
\def\tdk{{\ensuremath{\tilde k}}}

\def\tdm{{\ensuremath{\tilde m}}}
\def\tdn{{\ensuremath{\tilde n}}}

\def\tdq{{\ensuremath{\tilde q}}}

\def\bbb{{\ensuremath{\mathbf b}}}

\def\bbd{{\ensuremath{\mathbf d}}}

\def\bbq{{\ensuremath{\mathbf q}}}

\def\bbu{{\ensuremath{\mathbf u}}}

\def\bb0{{\ensuremath{\mathbf 0}}}
\def\tbq{{\tilde{\ensuremath{\mathbf q}} }}

\def \QED {$\square$}

\newtheorem{proposition}{\hspace{0pt}\bf Proposition}

\newtheorem{theorem}{\hspace{0pt}\bf Theorem}
\newtheorem{corollary}{\hspace{0pt}\bf Corollary}
\newtheorem{fact}{\hspace{0pt}\bf Fact}
\newtheorem{remark}{\hspace{0pt}\bf Remark}

\newtheorem{definition}{\hspace{0pt}\bf Definition}

\newcommand{\norm}{ {\|\cdot\|_p} }
\newcommand{\Nnorm}{ {\ccalN, p }}
\newcommand{\Dnorm}{ {\ccalD, p }}
\newcommand{\Pnorm}{ {\ccalP, p }}

\newcommand{\tcQ}{\tilde\ccalQ}

\newcommand{\conv}{\text{conv}}

\newcommand \PL    [1] {\ccalP_{#1} \ccalL}

\begin{document}

\title{Persistent Homology Lower Bounds on \\ High Order Network Distances}

\author{Weiyu Huang and Alejandro Ribeiro
\thanks{Work supported by NSF CCF-1217963 and AFOSR MURI FA9550-10-1-0567. The authors are with the Department of Electrical and Systems Engineering, University of Pennsylvania, 200 South 33rd Street, Philadelphia, PA 19104. Email: \{whuang, aribeiro\}@seas.upenn.edu. Preliminary results appeared in \cite{Huang15b, Huang16}.}}

\maketitle

\begin{abstract}
High order networks are weighted hypergraphs collecting relationships between elements of tuples, not necessarily pairs. Valid metric distances between high order networks have been defined but they are difficult to compute when the number of nodes is large. The goal here is to find tractable approximations of these network distances. The paper does so by mapping high order networks to filtrations of simplicial complexes and showing that the distance between networks can be lower bounded by the difference between the homological features of their respective filtrations. Practical implications are explored by classifying weighted pairwise networks constructed from different generative processes and by comparing the coauthorship networks of engineering and mathematics academic journals. The persistent homology methods succeed in identifying different generative models, in discriminating engineering and mathematics communities, as well as in differentiating engineering communities with different research interests. \end{abstract}


\IEEEpeerreviewmaketitle

%
\section{Introduction}\label{sec_introduction}

High order networks describe relationships between elements of tuples that go beyond the pairwise relationship that are more often considered \cite{Ahuja93, Wasserman94}. High order networks are important in applications in which relationships between triplets, quadruplets, and generic $n$-tuples are important. For example, coauthorship networks can be constructed by detailing the number of joint publications by pairs of scholars, but adding information on publications of individual authors and joint publications by specific triplets of authors generates a more accurate description. The relevance of expressing tuple relationships other than pairwise has been noted in various domains, including sensor networks \cite{Ghrist05, Chintakunta10}, cognitive learning \cite{Zhang08}, wireless networking \cite{Ren12}, image ranking \cite{Xu12}, object recognition \cite{Gao12}, and social networks \cite{Wilkerson13}.

The goal of this paper is to develop efficient methods to quantify differences between high order networks of possibly different sizes. Ideally, differences should be quantified using distances derived from an underlying metric in the space of high order networks. However, while distances are not difficult to define, they can be intractable to compute. This difficulty has motivated the use of feature comparisons in which the difference between specific properties of the network is used as a more manageable alternative. Examples of feature comparisons are clustering coefficients \cite{Wang12}, neighborhood topologies \cite{Singh08}, betweenness \cite{Peng10}, motifs \cite {Choobdar12}, wavelets \cite{Yong10}, and graphlet degree distributions \cite{Przulj07, Milenkovic08}. While computationally simpler, the use of features is application dependent, utilizes only a small portion of the information conveyed by the networks, and may yield conflicting comparative judgements because the triangle inequality is not necessarily valid. A proper distance between pairwise and high order networks overcomes these drawbacks. This is the motivation for the definition of network metrics that generalize the Gromov-Hausdorff distance between metric spaces \cite{Gromov07, Memoli08} to pairwise \cite{Carlsson14} and high order networks \cite{Huang15}.

The metric distances between high order networks defined in \cite{Huang15} have been applied to compare networks with small number of nodes and have succeeded in identifying collaboration patterns of coauthorship networks. However, because they have to consider all possible node correspondences (Definition \ref{dfn_correspondence}), network distances are difficult to compute when the number of nodes in the networks is large. The goal of this paper is to develop network discrimination methods that are computable in networks with large numbers of nodes. These discrimination methods are constructed by drawing a parallel between high order networks and algebraic topology filtrations. Homological features of filtrations are then used to compare high order networks and shown to provide {\it lower bounds} for the actual network distances. Although distance lower bounds suffer from some of the same problems associated with feature comparisons, they nonetheless have important properties. Among them, we know that a large lower bound entails a large distance and that we can use lower bounds to estimate distance intervals because upper bounds are easy to determine.

The paper begins with a review of high order networks and high order network distances (Section \ref{sec_high_order_network} and \cite{Huang15}). We then introduce concepts from algebraic topology \cite{Edelsbrunner00, Zomorodian05} and discuss the representation of high order networks in terms of filtrations (Section \ref{sec_filtration_persistence}). Specifically, we revisit the definition of a homological feature as a formalization of the notion of a hole without an interior in an unweighted network and the definition of a filtration as a nested collection of unweighted networks indexed by a time parameter. The most important observation in this preliminary discussion is that a class of high order networks can be represented as a filtration that starts empty and to which nodes, edges, and high order tuples are added at a time given by their weight (Section \ref{sec_filtrations}). We then introduce persistence diagrams to monitor the appearance and disappearance of homological features from the unweighted networks that compose a filtration (Section \ref{sec_persistence}). 

We propose to compare networks by considering their respective filtrations and using the difference between the persistence diagrams of the respective filtrations as a proxy for their distance. (Section \ref{sec_persistence_lower_bound}). This proposed methodology is substantiated by the fact that we can lower bound the computationally intractable distance between two high order networks with a tractable distance between their respective persistence diagrams (Theorems \ref{thm_DN} and \ref{thm_DN_k} in Section \ref{sec_persistence_lower_bound}). These lower bounds are tight such that there are examples in which the lower bound and the actual distances coincide. Since persistent homologies can be computed efficiently for large networks (Section \ref{sec_numerical} and \cite{Mischaikow13, Harker14}), we can use these lower bounds in, e.g., network classification problems. We do so for artificial networks created with different models -- random networks, Gaussian kernel proximity networks, and Euclidean feature networks -- and for coauthorship networks constructed from the publications of a number of journals from engineering and mathematics communities. The proposed methods succeed in distinguishing networks with different generative models (Section \ref{sec_synthetic}) and are also effective in discriminating engineering journals from mathematics journals (Section \ref{sec_compare_coauthorship}). We also attempt a more challenging classification problem of three different engineering communities where we achieve a moderate success (Section \ref{sec_engineering_different}). We close with concluding remarks (Section \ref{sec_conclusion}). 

%
\section{High Order Networks}\label{sec_high_order_network}

A network $N_X^K$ of order $K$ over the node set $X$ is defined as a collection of $K+1$ normalized relationship functions $\{r_X^k : X^{k+1} \rightarrow [0, 1]\}_{k = 0}^{K}$ from the space $X^{k+1}$ of $(k+1)$-tuples to the closed unit interval $[0, 1]$. For tuples $x_{0:k} := x_0, x_1, \dots, x_k$, the values $r_X^k(x_{0:k})$ of their $k$-order relationship functions are intended to represent a measure of similarity or dissimilarity for members of the group. We restrict attention to symmetric networks defined as those in which the relationship function depends on the elements of the tuple but not on ordering nor repetition. This implies, e.g., that $r_X^2(x, x, x') = r_X^1(x, x')$ because $x$ is repeated in the tuple $x,x,x'$ and that  $r_X^2(x, x', x'') = r_X^2(x', x, x'') $ because the tuples $x, x', x''$ and $x', x, x''$ contain the same elements in different order. Observe that the network $N_X^K$ can be considered as a weighted complete hypergraph \cite{Berge76} whose weights for all possible hyperedges are defined. Since relationships are not explicitly defined for all hyperedges in many cases, we adopt the convention that the undefined hyperedge relationships take the value $0$ or $1$, depending on whether the relationships encode dissimilarities or similarities (see Section \ref{sec_dissimilarity_network} and Section \ref{sec_proximity_network}). The value of $K$ is no greater than $2$ in most applications. The set of all high order networks of order $K$ is denoted as $\ccalN^K$.

Two networks $N_X^K$ and $N_Y^K$ are said $k$-isomorphic if there exists a bijection $\pi: X \rightarrow Y$ such that for all $x_{0:k} \in X^{k+1}$ we have
\begin{align}\label{eqn_N_isomorphic}
      r_Y^k (\pi(x_{0:k})) = r_X^k(x_{0:k}),
\end{align}
where we use the shorthand notation $r_Y^k (\pi(x_{0:k})):= r_Y^k (\pi(x_0), \pi(x_1), \dots, \pi(x_k))$. Since the map $\pi$ is bijective, \eqref{eqn_N_isomorphic} can only be satisfied when $X$ is a permutation of $Y$. When networks $N_X^K$ and $N_Y^K$ are $k$-isomorphic we write $N_X^K \cong_k N_Y^K$. The space of $K$-order networks where $k$-isomorphic networks are represented by the same element is termed the set of $K$-networks modulo $k$-isomorphism and denoted by $\ccalN^K \mod \cong_k$. For each $0 \le k \le K$, the space $\ccalN^K \mod \cong_k$ can be endowed with a pseudometric \cite{Huang15}. The definition of this distance requires introducing the notion of correspondence \cite[Def. 7.3.17]{Burago01}:

%
\begin{definition}\label{dfn_correspondence}
A correspondence between two sets $X$ and $Y$ is a subset $C \subseteq X \times Y$ such that $\forall~x \in X$, there exists $y \in Y$ such that $(x,y) \in C$ and $\forall~ y \in Y$ there exists $x \in X$ such that $(x,y) \in C$. The set of all correspondences is denoted as $\ccalC(X,Y)$.
\end{definition}

%
A correspondence in the sense of Definition \ref{dfn_correspondence} connects node sets $X$ and $Y$ so that every element of each set has at least one correspondent in the other set. We can now define the distance between two networks by selecting the correspondence that makes them the most similar as we formally define next.

%
\begin{definition}\label{dfn_d_N}
Given networks $N_X^K$ and $N_Y^K$, a correspondence $C$ between the node sets $X$ and $Y$, and an integer $0 \le k \le K$ define the $k$-order network difference with respect to $C$ as 
\begin{align}\label{eqn_d_N_prelim} 
   \Gamma_{X,Y}^k (C)
      := \max_{(x_{0:k}, y_{0:k}) \in C}  \
            \left| r_X^k(x_{0:k}) - r_Y^k (y_{0:k}) \right|,
\end{align}
where $(x_{0:k}, y_{0:k})$ stands for $(x_0, y_0),$  $(x_1, y_1), \dots, (x_k, y_k)$. The $k$-order network distance between $N_X^K$ and $N_Y^K$ is defined as
\begin{align}\label{eqn_d_N}
   d_\ccalN^k (N_X^K, N_Y^K) := \min_{C \in \ccalC(X,Y)} \
   \left\{ \Gamma_{X,Y}^k (C) \right\}. \end{align}\end{definition}

%
For a given correspondence $C \in \ccalC(X,Y)$ the network difference $\Gamma_{X,Y}^k(C)$ selects the maximum distance difference $|r_X^k(x_{0:k}) - r_Y^k (y_{0:k})|$ among all pairs of correspondents -- we compare $r_X^k(x_{0:k}) $ with $r_Y^k (y_{0:k})$ when all the points $x_l \in x_{0:k}$ and $y_l \in y_{0:k}$ are correspondents. The distance in \eqref{eqn_d_N} is defined by selecting the correspondence that minimizes these maximal differences. The distance in Definition \ref{dfn_d_N} is a pseudometric in $\ccalN^K \mod \cong_k$ \cite{Huang15}. Observe that since correspondences may be between networks with different number of elements, $d_\ccalN^k(N_X^K, N_Y^K)$ is defined when the node cardinalities $|X|$ and $|Y|$ are different. 

The distances in \eqref{dfn_d_N} compare relationship functions of a given order $k$. To compare all orders $0 \le k \le K$ we begin by defining two $K$-order networks $N_X^K$ and $N_Y^K$ as isomorphic if there exists a bijection $\pi: X \rightarrow Y$ such that \eqref{eqn_N_isomorphic} holds for all $0 \le k \le K$ and $x_{0:k} \in X^{k+1}$. When networks $N_X^K$ and $N_Y^K$ are isomorphic we write $N_X^K \cong N_Y^K$. The space of $K$-order networks modulo isomorphism is denoted as $\ccalN^K \mod \cong$. A family of pseudometrics measuring the difference over all order functions can be endowed to $\ccalN^K \mod \cong$ as we formally state next.  

%
\begin{definition}\label{dfn_d_N_norm}
Given networks $N_X^K$ and $N_Y^K$, a correspondence $C$ between the node sets $X$ and $Y$, and some vector $p$-norm $\norm$, define the network difference with respect to $C$ as 
\begin{align}\label{eqn_d_N_norm_prelim} 
    \left\| \bold \Gamma^K_{X,Y} (C) \right\|_p \!:= \!
    \left\| \!\big( \Gamma_{X,Y}^0 (C), \Gamma_{X,Y}^1 (C), \dots, 
          \Gamma_{X,Y}^K (C) \big)^T\! \right\|_p\!,\!
\end{align}
where $\Gamma_{X,Y}^k (C)$ is the $k$-order difference as per $C$ defined in \eqref{eqn_d_N_prelim}. The $p$-norm network distance between $N_X^K$ and $N_Y^K$ is defined as
\begin{align}\label{eqn_d_N_norm}
     d_\Nnorm (N_X^K, N_Y^K) := \min_{C \in \ccalC(X,Y)} \
           \left\{ \big\| \bold \Gamma^K_{X,Y} (C) \big\|_p \right\}.
\end{align} \end{definition}

%
In Definition \ref{dfn_d_N_norm} we compare relationship functions of all orders no larger than $K$. The norm $\|\bold \Gamma^K_{X,Y}(C)\|_p$ is assigned as the difference between $N_X^K$ and $N_Y^K$ measured by the correspondence $C$. The distance $d_\Nnorm (N_X^K, N_Y^K)$ is then the minimum of these differences achieved by some correspondence. 

The $k$-order function $r_X^k$ of a network $N^K_X$ does not impose constraints on the $l$-order function $r_X^l$ of the same network. In practical situations, however, it is common that adding nodes to a tuple results in increasing or decreasing relationships for the extended tuple. This motivates dissimilarity networks, which we do next, and proximity networks, which we discuss in Section \ref{sec_proximity_network}.

%
\begin{figure}[t]
\centerline{\def \thisplotscale {0.45}
\def \unit {\thisplotscale cm}

\pgfdeclarelayer{background}
\pgfdeclarelayer{foreground}
\pgfsetlayers{background,foreground}

\begin{tikzpicture}[-stealth, shorten >=0, x = 1.05*\unit, y=0.4*\unit, font=\scriptsize]

    \begin{pgfonlayer}{foreground}
       \node [blue vertex, minimum width = 1.2 * \unit, minimum height = 1.2 * \unit] at ( 2, 8) (A) {$A$}; \node at ( 2, 9.8) {$0$};
       \node [blue vertex, minimum width = (1.2 - 1/18) * \unit, minimum height = (1.2 - 1/18) * \unit] at ( 8, 8) (B) {$B$}; \node at ( 8, 9.8) {$1/9$};  
       \node [blue vertex, minimum width = (1.2 - 5/18) * \unit, minimum height = (1.2 - 5/18) * \unit] at ( -1.5, 0) (C) {$C$}; \node at (-3, 0) {$5/9$};  
       \node [blue vertex, minimum width = (1.2 - 3/18) * \unit, minimum height = (1.2 - 3/18) * \unit] at (11.5, 0) (D) {$D$}; \node at (13, 0) {$3/9$};  
       \path [-, above, line width = 21/9 pt] (A) edge node {{$ 2/9$}} (B);	
       \path [-, left, line width = 12/9 pt] (A) edge node {{$5/9 + \epsilon$}} (C);	
       \path [-, below, pos = 0.85, line width = 6/9 pt] (B) edge node [right] {{$7/9$}} (C);	
       \path [-, right, line width = 15/9 pt] (B) edge node [right]{{$4/9$}} (D);
       \path [-, below, pos = 0.85, line width = 15/9 pt] (A) edge node [left] {{$4/9$}} (D);
    \end{pgfonlayer}

    \begin{pgfonlayer}{background}    
      \filldraw [ultra thin, fill = orange!90!cyan, fill opacity = 0.2] 
                (A.center) --(B.center) --(C.center) -- cycle;
      \node at (2.3, 5) {$8/9$};  
      \filldraw [ultra thin, fill = orange!40!cyan, fill opacity = 0.2] 
                (A.center) --(B.center) --(D.center) -- cycle;
    \node at (7.7, 5) {$4/9 + \epsilon$};  
    \end{pgfonlayer}
        
\end{tikzpicture}\vspace{-3mm} }
\caption{Temporal dynamics for the formation of a research community. The $k$-order relationship in this 2-order dissimilarity network [cf. Definition \ref{dfn_dissimilarity_network}] denotes the normalized time instant when members of a $(k+1)$-tuple write their first joint paper. E.g., $A$ writes her first paper at time $0$, and coauthors with $B$, $D$, $C$ at times $2/9$, $4/9$, $5/9$, respectively. The $\epsilon$ in the relationship function is a technical modification to differentiate relationship between two authors from that of a single author. Author $A$ also writes jointly with $B$ and $D$ at time $4/9$.\vspace{-3mm}
}
\label{fig_dissimilarity_network_example}
\end{figure}
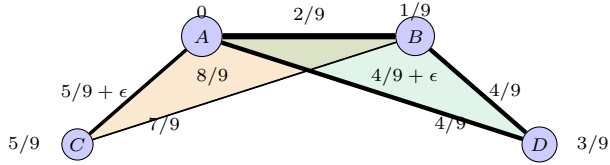

%
\subsection{Dissimilarity Networks}\label{sec_dissimilarity_network}

In dissimilarity networks, $r_X^k(x_{0:k})$ encodes a level of dissimilarity between elements. In this scenario it is reasonable that adding elements to a tuple makes the group more dissimilar and results in a higher value in the relationship function. This restriction makes up the formal definition that we introduce next. 

%
\begin{definition}\label{dfn_dissimilarity_network}
We say that the $K$-order network $D_X^K = \left(X, r_X^0, \dots, r_X^K\right)$ is a dissimilarity network if the order increasing property holds, i.e. for any order $1 \le k \le K$ and tuples $x_{0:k} \in X^{k+1}$ we have
\begin{align}\label{eqn_dfn_order_increasing}
   r_X^k(x_{0:k}) \geq r_X^{k-1}(x_{0:k-1}),
\end{align}
and the inequality \eqref{eqn_dfn_order_increasing} equalizes if and only if all the elements in $x_{0:k}$ also appear in $x_{0:k-1}$. Denote the set of all dissimilarity networks of order $K$ as $\ccalD^K$.
\end{definition}

%
As per Definition \ref{dfn_dissimilarity_network} dissimilarities $r_X^k(x_{0:k})$ grow as we add elements to a tuple. Notice that due to symmetry, a relationship as in \eqref{eqn_dfn_order_increasing} holds if we remove an arbitrary node from the tuple $x_{0:k}$, not necessarily the last. We further emphasize that Definition \ref{dfn_dissimilarity_network} includes the condition of having the inequality in \eqref{eqn_dfn_order_increasing} be strict unless all the elements of $x_{0:k}$ appear in $x_{0:k-1}$. This implies that to have $r_X^k(x_{0:k}) = r_X^{k-1}(x_{0:k-1})$ it must be that at least one element of $x_{0:k}$ is repeated in the tuple. Conversely, when we add a {\it distinct} element to a tuple, the dissimilarity must increase, i.e, $r_X^k(x_{0:k-1}, x_k) > r_X^{k-1}(x_{0:k-1})$ if $x_k$ is not a repetition of an element of $x_{0:k-1}$. This condition is technical and of little consequence for the results presented later --  see Remark \ref{rmk_epsilon_summands}. 

To see why the order increasing property is reasonable, consider a network describing the temporal dynamics of the formation of a research community -- see Figure~\ref{fig_dissimilarity_network_example}. The relationship function in this network marks the normalized time instant at which members of a given $(k+1)$-tuple write their first joint paper. For $k=0$, the zeroth order dissimilarities $r_X^0$ are the normalized time instants when authors publish their first paper. In Figure~\ref{fig_dissimilarity_network_example} authors $A$, $B$, $C$, and $D$ publish their first papers at times $0$, $1/9$, $5/9$, and $3/9$. The first order dissimilarities $r_X^1$ between pairs denote times at which nodes become coauthors. Since a coauthored paper is also a paper by each of the authors, it is certain that $r_X^1(x, x')\geq r_X^0(x)$. In Figure~\ref{fig_dissimilarity_network_example}, $A$ and $B$ become coauthors at time $2/9$, which occurs after they publish their respective first papers at times $0$ and $1/9$. Authors $A$ and $D$ become coauthors at time $4/9$ and $A$ and $C$ become coauthors at time $5/9$. Second order dissimilarities $r_X^2$ denote the time at which a paper is coauthored by a triplet. Since a paper can't be coauthored by three people without being at the same time coauthored by each of the three possible pairs of authors we must have $r_X^2(x,x',x'')\geq r_X^1(x,x')$. In Figure~\ref{fig_dissimilarity_network_example}, authors $A$, $B$, and $D$ publish a joint paper at time $4/9$, which is no smaller than the pairwise coauthorship times between each two of the individual authors. Undefined hyperedge relationships in dissimilarity networks take the value $1$ indicating the maximum dissimilarities for the tuple. An additive constant $\epsilon$ is used to make $r_X^1(A, C)$ strictly larger than $r_X^0(C)$ and $r_X^2(A, B, D)$ strictly larger than $r_X^1(A, D)$ and $r_X^1(B, D)$. This is needed because of the requirement in Definition \ref{dfn_dissimilarity_network} that the equality in \eqref{eqn_dfn_order_increasing} be achieved only if the elements in $x_{0:k}$ are all elements of $x_{0:k-1}$.

When specialized to dissimilarity networks, the network distances in definitions \ref{dfn_d_N} 
and \ref{dfn_d_N_norm} are termed the $k$-order dissimilarity network distance $d_\ccalD^k$ and the $p$-norm dissimilarity network distance $d_\Dnorm$. The restriction to dissimilarities makes the distances well-defined {\it metrics} in their respective spaces \cite{Huang15}. Although these metrics provide well-founded methods to compare high order networks, the search for the optimal correspondences in \eqref{eqn_d_N}  and \eqref{eqn_d_N_norm} is combinatorial. This makes it impossible to evaluate distances when the number of nodes in the networks is large and motivates the development of tractable lower bounds. These bounds are obtained by relating dissimilarity networks to filtrations of simplicial complexes \cite{Edelsbrunner00, Zomorodian05} as we explain next.

%
\begin{figure}[t]
\centerline{\def \thisplotscale {0.45}
\def \unit {\thisplotscale cm}

\pgfdeclarelayer{back}
\pgfdeclarelayer{fore}
\pgfdeclarelayer{mid}
\pgfsetlayers{back,mid,fore}

\begin{tikzpicture}[-stealth,  shorten >=0, x = 0.9*\unit, y=0.8*\unit, font=\scriptsize]


    \begin{pgfonlayer}{fore}
    \draw[dotted,->] (-1.5,0) -- (1.5,0);
    \node [blue vertex] at (0,0) (v) {$a$};
    \node at (0, -4) {vertex $[a]$};	
    \node at (0, -5) {$0$-simplex};	
    \node [blue vertex] at (6,-1) (e1) {$a$};
    \node [blue vertex] at (4,1) (e2) {$b$}; 
    \draw[dotted,->] (3.5,-1) -- (7,-1);
    \draw[dotted,->] (4, -1.5) -- (4, 2);
    \node at (5, -4) {edge $[a, b]$};	
    \node at (5, -5) {$1$-simplex};	 
    \node [blue vertex] at (12.5, -1) (tr1) {$a$};
    \node [blue vertex] at (10.5, 1) (tr2) {$b$}; 
    \node [blue vertex] at (10.5 - 1.414, -1 - 1.414) (tr3) {$c$}; 
    \draw[dotted,->] (10, -1) -- (13.5, -1);
    \draw[dotted,->] (10.5, -1.5) -- (10.5, 2);
    \draw[dotted,->] (10.5 + 2.12 / 6, -1 + 2.12 / 6) -- (10.5 - 2.12, -1 - 2.12);
    \node at (11, -4) {triangle $[ a,b, c]$};	
    \node at (11, -5) {$2$-simplex};	  
    \node [blue vertex] at (19.5, -1) (ti1) {$a$};
    \node [blue vertex] at (17.5, 1) (ti2) {$b$}; 
    \node [blue vertex] at (17.5 - 5.2 / 3, -1 + 1) (ti3) {$c$}; 
    \node [blue vertex] at (17.5 - 1, -1 - 5.2 / 3) (ti4) {$d$}; 
    \draw[dotted,->] (17.5 - 0.5, -1) -- (20.5, -1);
    \draw[dotted,->] (17.5, -1 - 0.5) -- (17.5, 2);
    \draw[dotted,->] (17.5 + 2.6 / 6, -1 - 1.5 / 6) -- (17.5 - 2.6, -1 + 1.5);
    \draw[dotted,->] (17.5 + 1.5 / 6, -1 + 2.6 / 6) -- (17.5 - 1.5, -1 -2.6);
    \node at (17.5, -4) {tetrahedron $[ a,b, c, d]$};
    \node at (17.5, -5) {$3$-simplex};		
    \end{pgfonlayer}
    
    \begin{pgfonlayer}{mid}
    \path[-, thick] (e1) edge (e2);	
    \path[-, thick] (tr1) edge (tr2);	
    \path[-, thick] (tr1) edge (tr3);	
    \path[-, thick] (tr2) edge (tr3);	
    \path[-, thick] (ti1) edge (ti2);	
    \path[-, thick] (ti1) edge (ti3);	
    \path[-, thick] (ti2) edge (ti3);	
    \path[-, thick] (ti1) edge (ti4);	
    \path[-, thick] (ti3) edge (ti4);	
    \path[dashed, -] (ti2) edge (ti4);	
    \end{pgfonlayer}
    
    \begin{pgfonlayer}{back}
    \filldraw[ultra thin, fill = cyan, fill opacity = 0.4] (tr1.center) --(tr2.center) --(tr3.center) -- cycle;
    \filldraw[ultra thin, fill = cyan, fill opacity = 0.4] (ti1.center) --(ti2.center) --(ti3.center) -- (ti4.center) -- cycle;
    \end{pgfonlayer}
     
\end{tikzpicture}\vspace{-3mm}}
\caption{Elementary $k$-simplices for $0 \le k \le 3$.\vspace{-4mm}
}
\label{fig_simplex}
\end{figure}
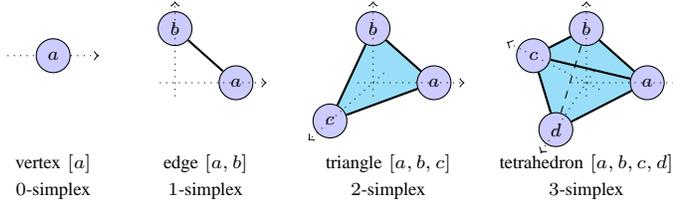

%
\section{Networks and Simplicial Complexes}\label{sec_filtration_persistence}

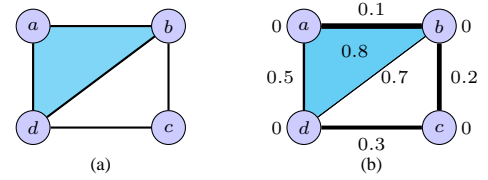
\begin{figure}[t]
\centerline{\def \thisplotscale {0.45}
\def \unit {\thisplotscale cm}

\pgfdeclarelayer{back}
\pgfdeclarelayer{fore}
\pgfdeclarelayer{mid}
\pgfsetlayers{back,mid,fore}

\begin{tikzpicture}[-stealth,  shorten >=0, x = 1*\unit, y=0.75*\unit, font=\scriptsize]

    
  \begin{pgfonlayer}{fore}
  
    \node [blue vertex] at (0, 0) (a) {$a$};
    \node [blue vertex] at (4, 0) (b) {$b$};
    \node [blue vertex] at (4, -4) (c) {$c$};
    \node [blue vertex] at (0, -4) (d) {$d$};
    
    \node at (2, -5.5) {(a)};
  \end{pgfonlayer}

  \begin{pgfonlayer}{mid}
    \path[-, thick] (a) edge (b);
    \path[-, thick] (b) edge (c);
    \path[-, thick] (c) edge (d);
    \path[-, thick] (d) edge (a);
    \path[-, thick] (d) edge (b);
  \end{pgfonlayer}
    
  \begin{pgfonlayer}{back}
    \filldraw[ultra thin, fill = cyan, fill opacity = 0.5] (a.center) --(b.center) --(d.center) -- cycle;
  \end{pgfonlayer}

    
\def \rightshift {8}

  \begin{pgfonlayer}{fore}
    
    \node [blue vertex] at (\rightshift + 0,0) (a) {$a$};
    \node at (\rightshift - 0.8, 0) {$0$};
    \node [blue vertex] at (\rightshift + 4, 0) (b) {$b$};
    \node at (\rightshift + 4+0.8, 0) {$0$};
    \node [blue vertex] at (\rightshift + 4, -4) (c) {$c$};
    \node at (\rightshift + 4+0.8, -4) {$0$};
    \node [blue vertex] at (\rightshift + 0, -4) (d) {$d$};
    \node at (\rightshift - 0.8, -4) {$0$};
     
    \node at (\rightshift + 2, -5.5) {(b)};
  \end{pgfonlayer}
    
  \begin{pgfonlayer}{mid}
    \path[-, above, line width = 2.1pt] (a) edge node {$0.1$}(b) ;
    \path[-, right, line width = 1.8pt] (b) edge node {$0.2$}(c) ;
    \path[-, below, line width = 1.5pt] (c) edge node {$0.3$}(d) ;
    \path[-, left, line width = 0.9pt] (d) edge node {$0.5$}(a) ;
    \path[-, right, line width = 0.3pt] (b) edge node {$0.7$}(d) ;
  \end{pgfonlayer}
    
  \begin{pgfonlayer}{back}
    \filldraw[ultra thin, fill = cyan, fill opacity = 0.6] (a.center) --(b.center) --(d.center) -- cycle;
    \node at (\rightshift + 1.5, -1) {$0.8$};
  \end{pgfonlayer}

\end{tikzpicture}\vspace{-3mm}}
\caption{(a) An example of simplicial complex $L$ which consists of four 0-simplices, five 1-simplices, and one 2-simplex. (b) A dissimilarity network can be represented as a simplicial complexes with weights. When the weight of a simplex denotes the time instant the simplex appears in the nested sequence of simplicial complexes, it yields a valid filtration $\ccalL$.\vspace{-3mm}
}
\label{fig_simplicial_complex}
\end{figure}

We introduce elemental notions of computational topology as they apply to the study of high order networks. Begin by defining a $k$-simplex $\phi = [x_{0:k}]$ as the convex hull of the set of points $x_{0:k}$ (see Figure \ref{fig_simplex}) and a simplicial complex $L$ as the collection of simplices such that for any simplex $[x_{0:k}]$, the convex hull of any subset of $x_{0:k}$ also belongs to $L$. Figure~\ref{fig_simplicial_complex} (a) exemplifies two triangles connected together as a simplicial complex of dimension $2$. It is a simplicial complex because for any of its simplices, say, $[a,b,c]$, the convex hulls of the subsets of $\{a,b,c\}$ -- which include the points $[a]$, $[b]$, and $[c]$ as well as the edges $[a,b]$, $[b,c]$, and $[a,c]$ -- all belong to the simplicial complex as well.

An important concept in simplicial complexes is that of a hole without interior. For the simplicial complex in Figure \ref{fig_simplicial_complex} (a), the area enclosed by $[b,d], [d,c], [c,b]$ is a hole without interior because the area is not filled. The area enclosed by $[a, b], [b, d], [d, a]$ is not because the interior is filled by the $2$-simplex $[a,b,d]$. Homologies are defined to formalize this intuition and rely on the definitions of chains, cycles, and boundaries. The $k$-{\it chain} $\Phi_k = \sum_i \beta_i \phi_i$ is a summation of $k$-simplices $\phi_i$ modulated by coefficients $\beta_i$ whose signs denote orientation. This definition is a generalization of the familiar definition of chains in graphs. E.g., in Figure \ref{fig_simplicial_complex}, $[a,b] + [b,d]$ is a $1$-chain  that we can equivalently represent as $[a,b] - [d,b]$. Further consider a given $k$-simplex $\phi = [x_{0:k}]$ and its border $(k-1)$-simplices defined as the ordered set of elemenets $[x_{0:\widehat l:k}]= \text{conv}\{ x_{0:k} \backslash x_l \}$, in which each of the elements is removed in order. The {\it boundary} $\partial_k \phi$ of the simplex $\phi$ is the chain formed by its borders using alternating orientations, 
\begin{align}\label{eqn_homology_boundary}
   \partial_k \phi = \sum_{l = 0}^k (-1)^l [x_{0:\widehat{l}:k}].
\end{align}
According to \eqref{eqn_homology_boundary}, the boundary $\partial_k \phi$ of a $k$-simplex is the collection of $(k-1)$-simplices. For the simplices in Figure \ref{fig_simplex}, the boundaries are $\partial_0[a] = 0, \partial_1 [a, b] = [b] - [a], \partial_2[a, b, c] = [b, c] - [a, c] + [a, b] $ and $\partial_3[a, b, c, d] = [b, c, d] - [a, c, d] + [a, b, d] - [a, b, c]$. 

Having defined chains and boundaries we consider a $k$-chain $\Phi_k = \sum_i\beta_i \phi_i$ and define the {\it chain boundary} $\partial_k \Phi_k$ as the summation of the boundaries of its component simplices,
\begin{equation}\label{eqn_homology_chain_boundary}
   \partial_k \Phi_k = \sum_i \beta_i (\partial_k \phi_i).
\end{equation}
As per this definition, the boundary of the chain $\Phi = [a,b] + [b,d]$ in Figure \ref{fig_simplicial_complex} is $\partial_k \Phi = [a] - [b] + [b] - [d] = [a] - [d]$, which matches our intuition of what the chain's boundary should be. We can now formally define a {\it $k$-cycle} $\Psi_k$ as a chain whose boundary is null, i.e., a $k$-chain for which $\partial_k \Psi = 0$. In Figure \ref{fig_simplicial_complex} the chain $\Psi = [a,b] + [b,d] + [d,a]$ is a cycle because its boundary is $\partial_k \Psi = [a] - [b] + [b] - [d] + [d] - [a] = 0$.

A $k$-cycle $\Psi_k$ is a $k$-{\it homological feature} if it cannot be represented as the boundary of a $(k+1)$-chain $\Phi_{k+1}$; otherwise $\Psi_k$ is a $k$-{\it boundary}. The group $\ccalH_k(L)$ of $k$-homological features of a simplicial complex $L$ represents all $k$-cycles that are not $k$-boundaries. I.e., a homological feature appears when a $k$-cycle $\Psi_k$ is not among the group of boundaries $ \partial_{k+1} \Phi_{k+1}$ of $(k+1)$-chains.

For the simplicial complex in Figure \ref{fig_simplicial_complex} (a), the chain $\Psi_1 = [a,b] + [b,d] + [d,a]$ is a $1$-cycle because $\partial_1\Psi_1 = 0$; simultaneously, $\Psi_1$ is also a $1$-boundary because $\partial_2 [a,b,d] = [b,d] - [a,d] + [a,b]$, which is identical to $\Psi_1$. Therefore, $\Psi_1$ does not represent a homological feature. On the other hand, the chain $\Psi_1' = [b,c] + [c,d] + [d,b]$ describes a homological feature because it is a cycle with $\partial_1 \Psi_1' = 0$, and not a boundary due to the fact that the $2$-simplex $[b,c,d]$ is absent from the complex.

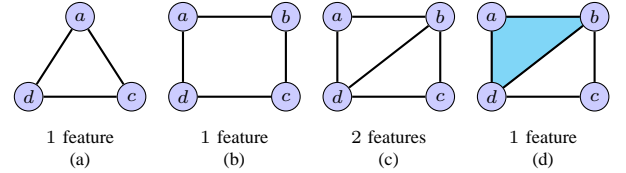
\begin{figure}[t]
\centerline{\def \thisplotscale {0.38}
\def \unit {\thisplotscale cm}

\pgfdeclarelayer{back}
\pgfdeclarelayer{fore}
\pgfdeclarelayer{mid}
\pgfsetlayers{back,mid,fore}
\begin{tikzpicture}[-stealth,  shorten >=0, x = 0.9*\unit, y=0.7*\unit, font=\scriptsize]

\def \downshift {1.2}

\begin{pgfonlayer}{fore}
    \node [blue vertex] at (3, 0 + 2) (a) {$a$};
    \node [blue vertex] at (3 + 2, 0 - 2) (b) {$c$};
    \node [blue vertex] at (3 - 2, 0 - 2) (c) {$d$};
\end{pgfonlayer}
    
\begin{pgfonlayer}{mid}
    \path[-, thick] (a) edge (b);
    \path[-, thick] (b) edge (c);
    \path[-, thick] (c) edge (a);
    \node at (3, -4) {$1$ feature};
    \node at (3, -4 - \downshift) {(a)};
\end{pgfonlayer}

\begin{pgfonlayer}{fore}
    \node [blue vertex] at (9 - 2, 0 + 2) (a) {$a$};
    \node [blue vertex] at (9 + 2, 0 + 2) (b) {$b$};
    \node [blue vertex] at (9 + 2, 0 - 2) (c) {$c$};
    \node [blue vertex] at (9 - 2, 0 - 2) (d) {$d$};
\end{pgfonlayer}
    
\begin{pgfonlayer}{mid}
    \path[-, thick] (a) edge (b);
    \path[-, thick] (b) edge (c);
    \path[-, thick] (c) edge (d);
    \path[-, thick] (d) edge (a);
    \node at (9, -4) {$1$ feature};
    \node at (9, -4 - \downshift) {(b)};
\end{pgfonlayer}
    
\begin{pgfonlayer}{fore}
    \node [blue vertex] at (15 - 2, 0 + 2) (a) {$a$};
    \node [blue vertex] at (15 + 2, 0 + 2) (b) {$b$};
    \node [blue vertex] at (15 + 2, 0 - 2) (c) {$c$};
    \node [blue vertex] at (15 - 2, 0 - 2) (d) {$d$};
\end{pgfonlayer}
    
\begin{pgfonlayer}{mid}
    \path[-, thick] (a) edge (b);
    \path[-, thick] (b) edge (c);
    \path[-, thick] (c) edge (d);
    \path[-, thick] (d) edge (a);
    \path[-, thick] (d) edge (b);
    \node at (15, -4) {$2$ features};
    \node at (15, -4 - \downshift) {(c)};
\end{pgfonlayer}
    
\begin{pgfonlayer}{back}
\end{pgfonlayer}

\begin{pgfonlayer}{fore}
    \node [blue vertex] at (21 - 2, 0 + 2) (a) {$a$};
    \node [blue vertex] at (21 + 2, 0 + 2) (b) {$b$};
    \node [blue vertex] at (21 + 2, 0 - 2) (c) {$c$};
    \node [blue vertex] at (21 - 2, 0 - 2) (d) {$d$};
\end{pgfonlayer}
    
\begin{pgfonlayer}{mid}
    \path[-, thick] (a) edge (b);
    \path[-, thick] (b) edge (c);
    \path[-, thick] (c) edge (d);
    \path[-, thick] (d) edge (a);
    \path[-, thick] (d) edge (b);
    \node at (21, -4) {$1$ feature};
    \node at (21, -4 - \downshift) {(d)};
\end{pgfonlayer}
    
\begin{pgfonlayer}{back}
    \filldraw[ultra thin, fill = cyan, fill opacity = 0.5] (a.center) --(b.center) --(d.center) -- cycle;
\end{pgfonlayer}

\end{tikzpicture} \vspace{-3mm}}
\caption{More examples of simplicial complexes and homologies. The number of $2$-homological features are described below each of the simplicial complex.}
\label{fig_homology}
\end{figure}

Other illustrative examples are shown in Figure \ref{fig_homology}. The complex (a) has one $1$-homological feature corresponding to the chain $[a,b] + [b,c] + [c,a]$ and the complex (b) has $1$ feature as well represented by $[a,b] + [b,c] + [c,d] + [d,a]$. Complex (c) has $2$ features that are represented by chains $[a,b] + [b,d] + [d,a]$ and $[b,c] + [c,d] + [d,b]$ but complex (d) has one feature because the chain $[a,b] + [b,d] + [d,a]$ is the boundary of the simplex $[a,b,d]$.

%

%

%
\subsection{Representation of High Order Networks as Filtrations}\label{sec_filtrations}

In the same manner in which a graph represents an unweighted network, a simplicial complex represents a high order unweighted network. To represent weighted high order networks we assign a weight to each simplex, but instead of thinking the network as a weighted simplicial complex we think of weights as parameters that indicates the time at which the simplex comes into existence. Formally, for parameters $\alpha \in [0,1]$ we define a filtration $\ccalL$ as a collection of simplicial complexes $L_\alpha$ such that for any ordered sequence $0 = \alpha_0 < \alpha_1 < \ldots < \alpha_m = 1$ it holds $\emptyset = L_{\alpha_0} \subseteq L_{\alpha_1} \subseteq  \ldots \subseteq L_{\alpha_m} = L$. The minimum time $\alpha$ at which a simplex becomes an element of $L_\alpha$ is the birth time of the simplex.

Given a dissimilarity network $D_X^K$, we construct a filtration $\ccalL(D_X^K)$ by assigning the appearing time of simplex $[x_{0:k}]$ as the relationship $r_X^k(x_{0:k})$ of the corresponding tuple, 
\begin{equation}\label{eqn_filtration_def}
   [x_{0:k}] \in L_\alpha \iff r_X^k(x_{0:k}) \leq \alpha .
\end{equation}
We show next that \eqref{eqn_filtration_def} defines a valid filtration.

%
\begin{proposition}\label{prop_filtration}
The filtration $\ccalL(D_X^K)$ with elements $L_\alpha$ induced from a given dissimilarity network $D_X^K$ through the relationship in \eqref{eqn_filtration_def} is a well defined filtration.
\end{proposition}

%
\begin{myproof}\label{proof_filtration}
For a given $\alpha$, \eqref{eqn_filtration_def} defines a set of complexes $L_\alpha$. It is clear that the collection of sets $L_\alpha$ is nested, i.e., that $\emptyset = L_{\alpha_0} \subseteq  \ldots \subseteq L_{\alpha_m} = L$ holds for any set of birth times $0 = \alpha_0 < \ldots < \alpha_m = 1$. To prove the statement we need to show that the set of complexes $L_\alpha$ is a valid simplicial complex. To show this it suffices to verify that for any $\alpha$, all faces of each simplex in $L_\alpha$ also appear no later than $\alpha$. Suppose simplex $[x_{0:k}]$ appears before time $\alpha$ for some chosen $\alpha$, then it must be true that $r_X^k(x_{0:k}) \leq \alpha$. For any faces of $[x_{0:k}]$, say $[x_{0:\tdk}]$ with $\tdk < k$, the order increasing property implies $r_X^\tdk(x_{0:\tdk}) \leq r_X^k(x_{0:k})$. Therefore,
\begin{align}\label{eqn_proof_filtration_final}
    r_X^\tdk(x_{0:\tdk}) \leq r_X^k(x_{0:k}) \leq \alpha,
\end{align}
which shows that the face $[x_{0:\tdk}]$ appears before time or on time $\alpha$. This means that $L_\alpha$ is a valid simplicial complex.
\end{myproof}

%
\begin{figure}[t]
\centerline{\def \thisplotscale {0.4}
\def \unit {\thisplotscale cm}

\pgfdeclarelayer{back}
\pgfdeclarelayer{fore}
\pgfdeclarelayer{mid}
\pgfsetlayers{back,mid,fore}
\tikzstyle{nodeSmaller} = [blue vertex,
                    minimum height = 0.8*\unit, 
                    minimum width  = 0.8*\unit]

\begin{tikzpicture}[-stealth,  shorten >=0, x = 0.9*\unit, y=0.9*\unit, font=\scriptsize]
  
    
  \begin{pgfonlayer}{fore}
    
    \node [nodeSmaller] at (-7 - 1, -5 + 1) (a) {$a$};
    \node [nodeSmaller] at (-7 + 1, -5 + 1) (b) {$b$};
    \node [nodeSmaller] at (-7 + 1, -5 -1) (c) {$c$};
    \node [nodeSmaller] at (-7 -1, -5 -1) (d) {$d$};
     
  \end{pgfonlayer}
    
  \begin{pgfonlayer}{mid}
    \node at (-7, -5 -2.5) {$t = 0$};
  \end{pgfonlayer}
    
  \begin{pgfonlayer}{back}
  \end{pgfonlayer}
  
    
  \begin{pgfonlayer}{fore}
    
    \node [nodeSmaller] at (-3 - 1, -5 + 1) (a) {$a$};
    \node [nodeSmaller] at (-3 + 1, -5 + 1) (b) {$b$};
    \node [nodeSmaller] at (-3 + 1, -5 -1) (c) {$c$};
    \node [nodeSmaller] at (-3 -1, -5 -1) (d) {$d$};
     
  \end{pgfonlayer}
    
  \begin{pgfonlayer}{mid}
    \path[-, above, line width = 2.1pt] (a) edge node {$$} (b);
    \node at (-0.5, -5) {$\cdots$};
    \node at (-3, -5 -2.5) {$t = 0.1$};
  \end{pgfonlayer}
    
  \begin{pgfonlayer}{back}
  \end{pgfonlayer}
  
    
  \begin{pgfonlayer}{fore}
    
    \node [nodeSmaller] at (2 - 1, -5 + 1) (a) {$a$};
    \node [nodeSmaller] at (2 + 1, -5 + 1) (b) {$b$};
    \node [nodeSmaller] at (2 + 1, -5 -1) (c) {$c$};
    \node [nodeSmaller] at (2 -1, -5 -1) (d) {$d$};
     
  \end{pgfonlayer}
    
  \begin{pgfonlayer}{mid}
    \path[-, above, line width = 2.1pt] (a) edge node {} (b);
    \path[-, right, line width = 1.8pt] (b) edge node {} (c);
    \path[-, below, line width = 1.5pt] (c) edge node {} (d);
    \path[-, right, line width = 0.9pt] (d) edge node {} (a);
    \node at (2, -5 -2.5) {$t = 0.5$};
  \end{pgfonlayer}
    
  \begin{pgfonlayer}{back}
  \end{pgfonlayer}
        
    
  \begin{pgfonlayer}{fore}
    
    \node [nodeSmaller] at (6 - 1, -5 + 1) (a) {$a$};
    \node [nodeSmaller] at (6 + 1, -5 + 1) (b) {$b$};
    \node [nodeSmaller] at (6 + 1, -5 -1) (c) {$c$};
    \node [nodeSmaller] at (6 -1, -5 -1) (d) {$d$};
     
  \end{pgfonlayer}
    
  \begin{pgfonlayer}{mid}
    \path[-, above, line width = 2.1pt] (a) edge node {} (b);
    \path[-, right, line width = 1.8pt] (b) edge node {} (c);
    \path[-, below, line width = 1.5pt] (c) edge node {} (d);
    \path[-, right, line width = 0.9pt] (d) edge node {} (a);
    \path[-, right, line width = 0.3pt] (d) edge node {} (b);
    \node at (6, -5 -2.5) {$t = 0.7$};
  \end{pgfonlayer}
    
  \begin{pgfonlayer}{back}
  \end{pgfonlayer}
  
    
  \begin{pgfonlayer}{fore}
    
    \node [nodeSmaller] at (10 - 1, -5 + 1) (a) {$a$};
    \node [nodeSmaller] at (10 + 1, -5 + 1) (b) {$b$};
    \node [nodeSmaller] at (10 + 1, -5 -1) (c) {$c$};
    \node [nodeSmaller] at (10 -1, -5 -1) (d) {$d$};
     
  \end{pgfonlayer}
    
  \begin{pgfonlayer}{mid}
    \path[-, above, line width = 2.1pt] (a) edge node {} (b);
    \path[-, right, line width = 1.8pt] (b) edge node {} (c);
    \path[-, below, line width = 1.5pt] (c) edge node {} (d);
    \path[-, right, line width = 0.9pt] (d) edge node {} (a);
    \path[-, right, line width = 0.3pt] (d) edge node {} (b);
    \node at (10, -5 -2.5) {$t = 0.8$};
    \node at (12.5, -5) {$\cdots$};
    \node at (2.5, -10.1) {filtration};
    \draw[->] (-9, -9.5) -- (14, -9.5);
    \draw[-] (-7, -9.5) -- (-7, -9.2);
    \draw[-] (-3, -9.5) -- (-3, -9.2);
    \draw[-] (2, -9.5) -- (2, -9.2);
    \draw[-] (6, -9.5) -- (6, -9.2);
    \node at (-0.5, -9.3) {$\cdots$};
    \draw[-] (10, -9.5) -- (10, -9.2);
    \node at (12.5, -9.3) {$\cdots$};
  \end{pgfonlayer}
    
  \begin{pgfonlayer}{back}
    \filldraw[ultra thin, fill = cyan, fill opacity = 0.6] (a.center) --(b.center) --(d.center) -- cycle;  \end{pgfonlayer}
     	
\end{tikzpicture} }
\begin{minipage}[h]{0.038\textwidth}
~
\end{minipage}
\begin{minipage}[h]{0.0692\textwidth}
    	\centering
    	\includegraphics[trim=0.06cm 0.1cm 0.1cm 0.1cm, clip=true, width=1 \textwidth]
	{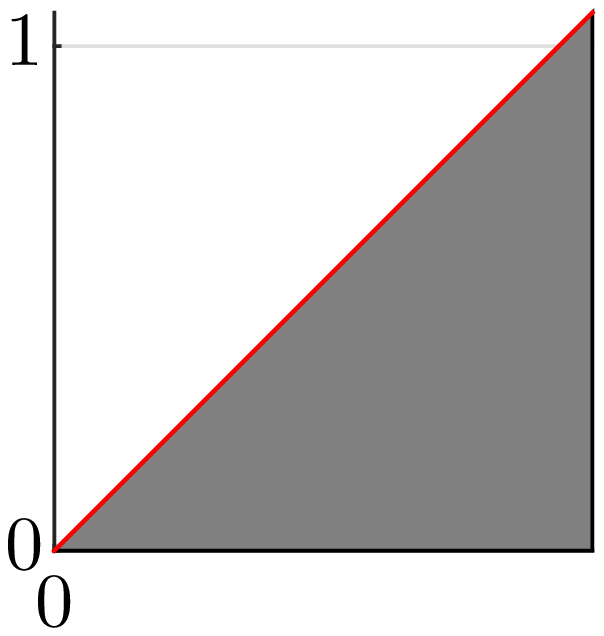}
\end{minipage}
\begin{minipage}[h]{0.0692\textwidth}
    	\centering
    	\includegraphics[trim=0.06cm 0.1cm 0.1cm 0.1cm, clip=true, width=1 \textwidth]
	{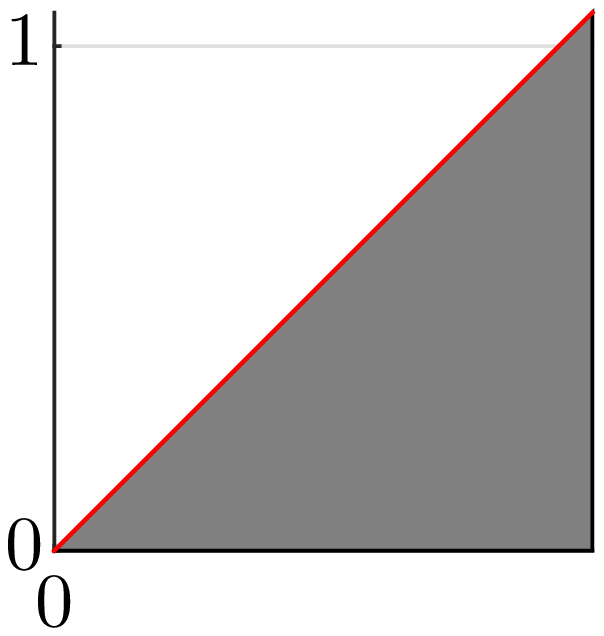}
\end{minipage}
\begin{minipage}[h]{0.014\textwidth}
{\tiny $\!\cdots$}
\end{minipage}
\begin{minipage}[h]{0.0692\textwidth}
    	\centering
    	\includegraphics[trim=0.06cm 0.1cm 0.1cm 0.1cm, clip=true, width=1 \textwidth]
	{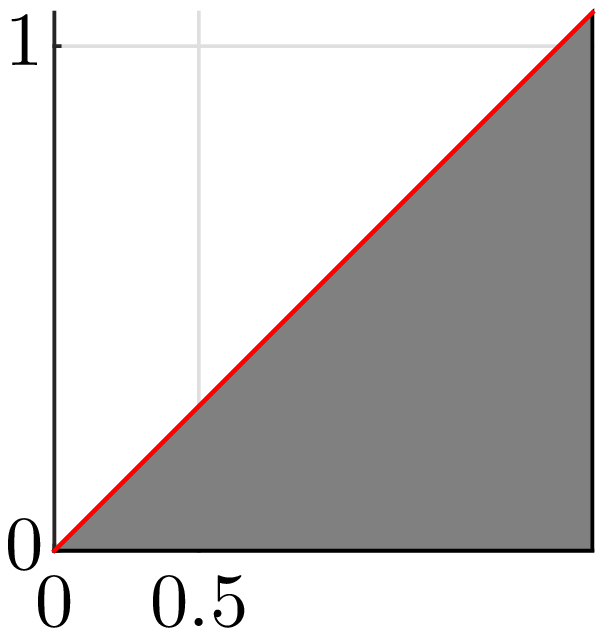}
\end{minipage}
\begin{minipage}[h]{0.0692\textwidth}
    	\centering
    	\includegraphics[trim=0.06cm 0.1cm 0.1cm 0.1cm, clip=true, width=1 \textwidth]
	{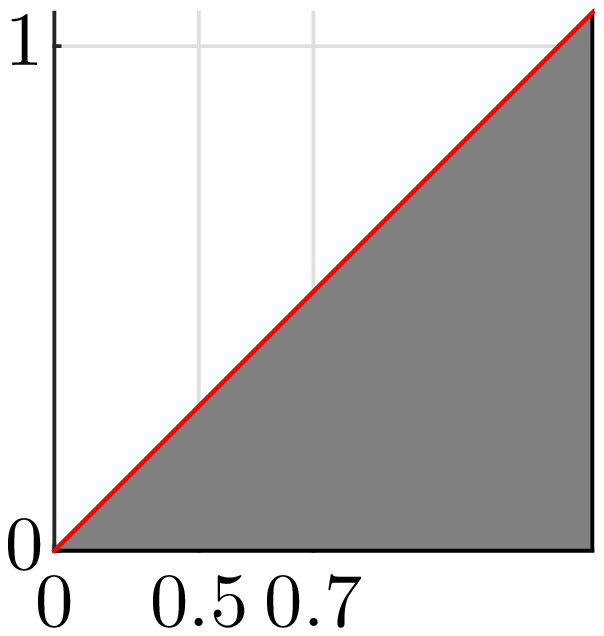}
\end{minipage}
\begin{minipage}[h]{0.0692\textwidth}
    	\centering
    	\includegraphics[trim=0.06cm 0.1cm 0.1cm 0.1cm, clip=true, width=1 \textwidth]
	{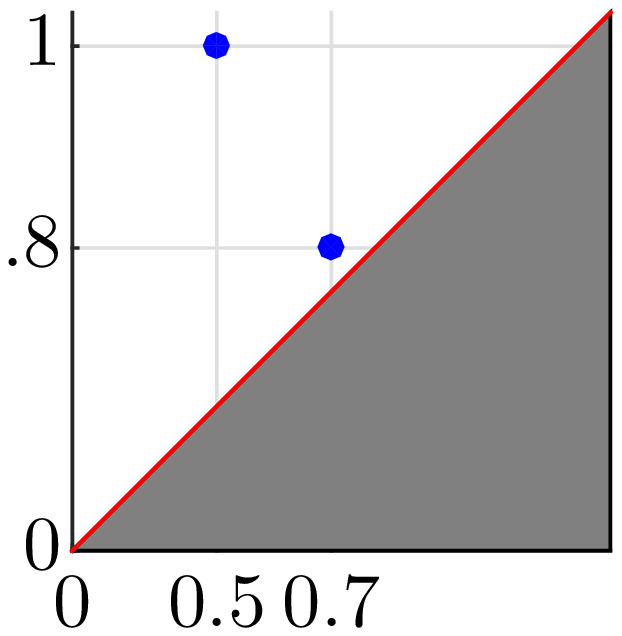}
\end{minipage}
\begin{minipage}[h]{0.014\textwidth}
{\tiny $\!\cdots$}
\end{minipage}
\caption{Top: filtration of nested simplicial complexes of the dissimilarity network exhibited in Figure \ref{fig_simplicial_complex} (b). The simplicial complex at each time instant are the collections of all vertices appearing before or on that time. Bottom: $1$-persistence diagrams describing $1$-persistent homologies for each of the simplicial complexes detailed in Top. The horizontal axis denotes the birth time of homological features and the vertical axis represents the death time.\vspace{0mm}}
\label{fig_filtration}
\centerline{\def \thisplotscale {0.35}
\def \unit {\thisplotscale cm}

\pgfdeclarelayer{back}
\pgfdeclarelayer{fore}
\pgfdeclarelayer{mid}
\pgfsetlayers{back,mid,fore}

\begin{tikzpicture}[-stealth,  shorten >=0, x = 1*\unit, y=0.9*\unit, font=\scriptsize]

\begin{pgfonlayer}{fore}
    \node [blue vertex] at (6, 0 + 2) (a) {$a$};
    \node [blue vertex] at (6 + 2, 0 - 2) (b) {$c$};
    \node [blue vertex] at (6 - 2, 0 - 2) (c) {$d$};
\end{pgfonlayer}
    
\begin{pgfonlayer}{mid}
    \path[-, right, line width = 1.8pt] (a) edge node {$0.2$} (b);
    \path[-, below, line width = 1.5pt] (b) edge node {$0.3$} (c);
    \path[-, left, line width = 0.9pt] (c) edge node {$0.5$} (a);
\end{pgfonlayer}

\begin{pgfonlayer}{fore}
    \node [blue vertex] at (12 - 2, 0 + 2) (a) {$a$};
    \node [blue vertex] at (12 + 2, 0 + 2) (b) {$b$};
    \node [blue vertex] at (12 + 2, 0 - 2) (c) {$c$};
    \node [blue vertex] at (12 - 2, 0 - 2) (d) {$d$};
\end{pgfonlayer}
    
\begin{pgfonlayer}{mid}
    \path[-, above, line width = 2.1pt] (a) edge node {$0.1$} (b);
    \path[-, right, line width = 1.8pt] (b) edge node {$0.2$} (c);
    \path[-, below, line width = 1.5pt] (c) edge node {$0.3$} (d);
    \path[-, right, line width = 0.9pt] (d) edge node {$0.5$} (a);
\end{pgfonlayer}
    
\begin{pgfonlayer}{fore}
    \node [blue vertex] at (18 - 2, 0 + 2) (a) {$a$};
    \node [blue vertex] at (18 + 2, 0 + 2) (b) {$b$};
    \node [blue vertex] at (18 + 2, 0 - 2) (c) {$c$};
    \node [blue vertex] at (18 - 2, 0 - 2) (d) {$d$};
\end{pgfonlayer}
    
\begin{pgfonlayer}{mid}
    \path[-, above, line width = 2.1pt] (a) edge node {$0.1$} (b);
    \path[-, right, line width = 1.8pt] (b) edge node {$0.2$} (c);
    \path[-, below, line width = 1.5pt] (c) edge node {$0.3$} (d);
    \path[-, right, line width = 0.9pt] (d) edge node {$0.5$} (a);
    \path[-, right, line width = 0.3pt] (d) edge node {$0.7$} (b);
\end{pgfonlayer}
    
\begin{pgfonlayer}{back}
\end{pgfonlayer}

\begin{pgfonlayer}{fore}
    \node [blue vertex] at (24 - 2, 0 + 2) (a) {$a$};
    \node [blue vertex] at (24 + 2, 0 + 2) (b) {$b$};
    \node [blue vertex] at (24 + 2, 0 - 2) (c) {$c$};
    \node [blue vertex] at (24 - 2, 0 - 2) (d) {$d$};
\end{pgfonlayer}
    
\begin{pgfonlayer}{mid}
    \path[-, above, line width = 2.1pt] (a) edge node {$0.1$} (b);
    \path[-, right, line width = 1.8pt] (b) edge node {$0.2$} (c);
    \path[-, below, line width = 1.5pt] (c) edge node {$0.3$} (d);
    \path[-, right, line width = 0.9pt] (d) edge node {$0.5$} (a);
    \path[-, right, line width = 0.3pt] (d) edge node {$0.7$} (b);
\end{pgfonlayer}
    
\begin{pgfonlayer}{back}
    \filldraw[ultra thin, fill = cyan, fill opacity = 0.5] (a.center) --(b.center) --(d.center) -- cycle;
    \node at (23.7, 1) {$0.8$};
\end{pgfonlayer}
     	
\end{tikzpicture} }
\begin{minipage}[h]{0.0218\textwidth}
~
\end{minipage}
\begin{minipage}[h]{0.104\textwidth}
    	\centering
    	\includegraphics[trim=0.06cm 0.1cm 0.1cm 0.1cm, clip=true, width=1 \textwidth]
	{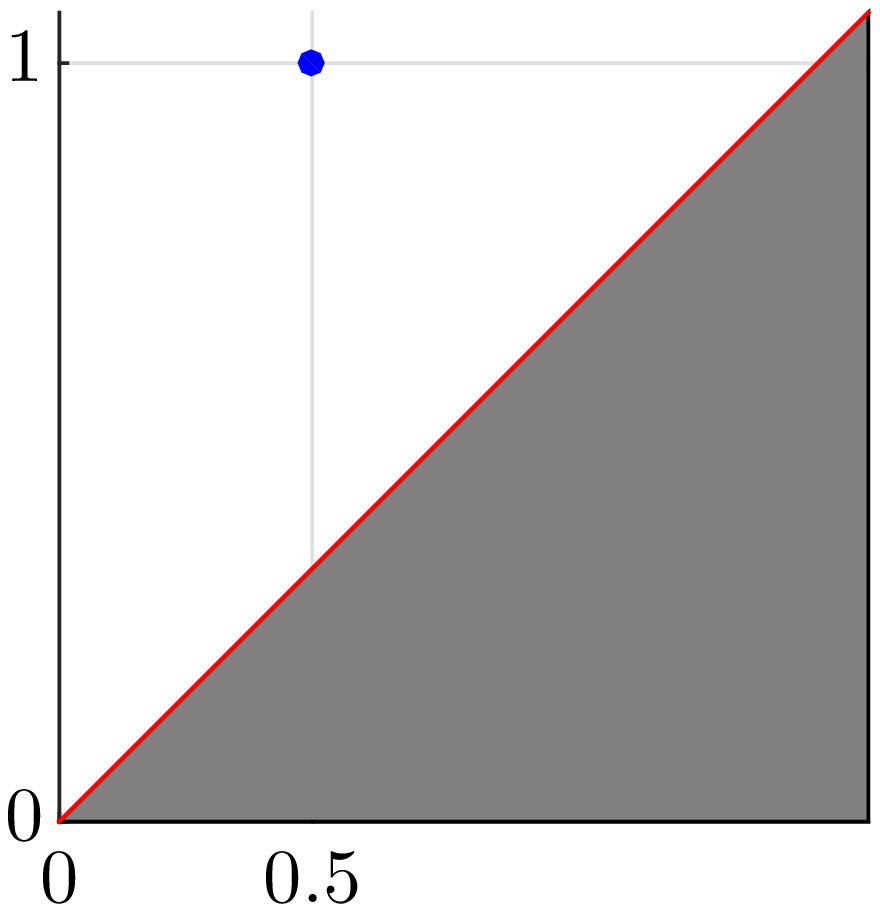}
	{\scriptsize (a)}
\end{minipage}
\begin{minipage}[h]{0.104\textwidth}
    	\centering
    	\includegraphics[trim=0.06cm 0.1cm 0.1cm 0.1cm, clip=true, width=1 \textwidth]
	{illus_1.eps}
	{\scriptsize (b)}
\end{minipage}
\begin{minipage}[h]{0.104\textwidth}
    	\centering
    	\includegraphics[trim=0.06cm 0.1cm 0.1cm 0.1cm, clip=true, width=1 \textwidth]
	{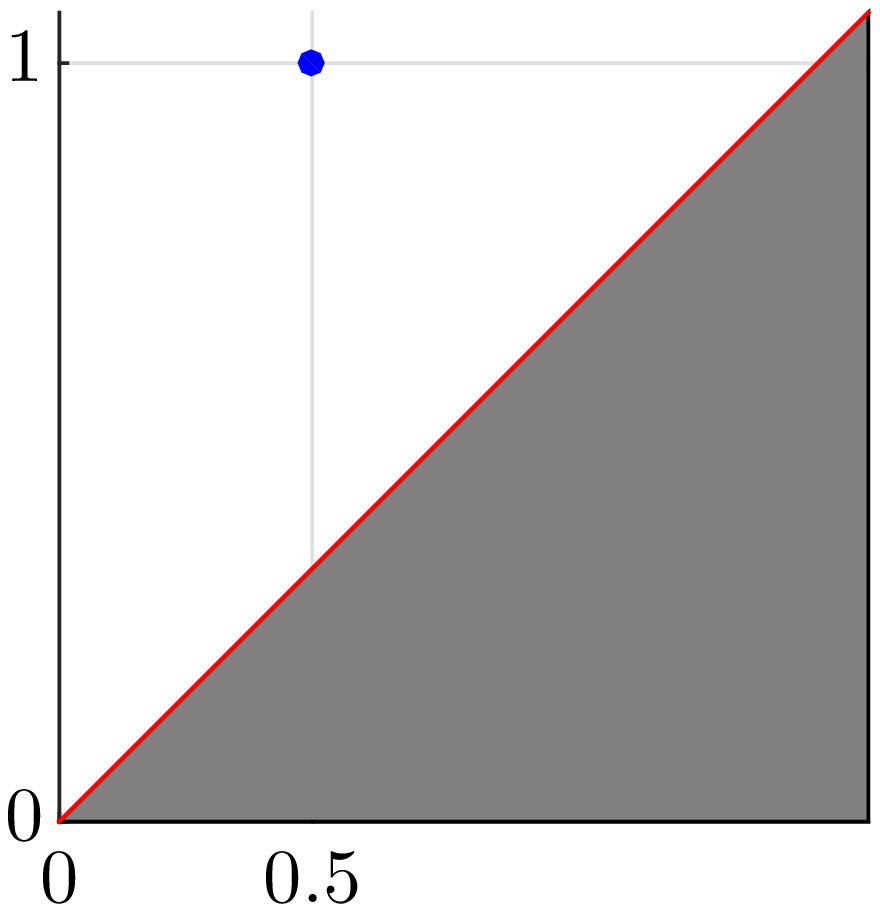}
	{\scriptsize (c)}
\end{minipage}
\begin{minipage}[h]{0.104\textwidth}
    	\centering
    	\includegraphics[trim=0.06cm 0.1cm 0.1cm 0.1cm, clip=true, width=1 \textwidth]
	{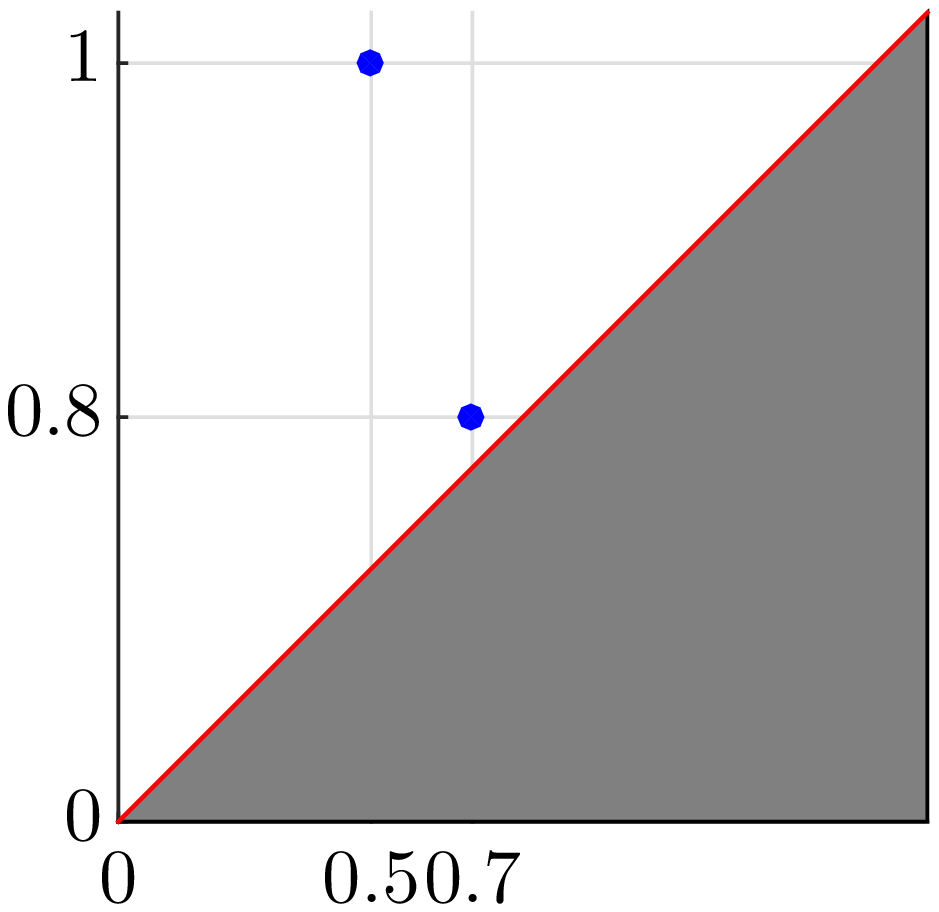}
	{\scriptsize (d)}
\end{minipage}
\caption{The $1$-persistence diagrams for the filtration induced by each of the dissimilarity networks. The maximum death time of any homological features is $1$ because undefined hyperedges take value $1$ implicitly. \vspace{-3mm}}
\label{fig_persistent_homology}
\end{figure}

%
We emphasize that filtrations can't be defined for an arbitrary high order networks because the order increasing property is necessary for the proof. For dissimilarity networks that do satisfy the order increasing property, Proposition \ref{prop_filtration} establishes the existence of an equivalent filtration.

Figure \ref{fig_filtration} (top) shows the construction of the filtration associated with the dissimilarity network in Figure \ref{fig_simplicial_complex} (b). At time $\alpha = 0$, there are four nodes because $r_X^0(a) = r_X^0(b) = r_X^0(c) = r_X^0(d) = 0$. At time $\alpha = 0.1$, the edge $[a,b]$ starts to appear in the simplicial complex because the relationship $r_X^1(a,b) = 0.1$. As time gradually increases, more simplices get included in the simplicial complex at each time instant. At the end of the filtration with $\alpha = 1$, all simplices in the original simplicial complex are involved.

%

{\subsection{Persistent Homologies and Persistence Diagrams}\label{sec_persistence}

As time advances in a filtration, holes and interiors appear and disappear. Persistent homologies examine when these homological features appear for the sequence of simplicial complexes in the filtration. Formally, given a filtration $\ccalL$, its $k$-dimensional persistence diagram $\PL k$ is a collection of points of the form $\bbq = [q_b, q_d]$ where $q_b$ and $q_d>q_b$ represent the birth and death time of a homological feature. I.e., the times $q_b$ represent resolutions $\alpha=q_b$ at which a feature is added to the group of $k$-homological features $\ccalH_k(L_{\alpha})=\ccalH_k(L_{q_b})$ and the times $q_d$ are resolutions $\alpha=q_d$ at which a feature is removed from the group of $k$-homological features $\ccalH_k(L_{\alpha})=\ccalH_k(L_{q_d})$.

The persistence diagram for the filtration in Figure \ref{fig_filtration}-(top) is shown in Figure \ref{fig_filtration}-(bottom). In this diagram the horizontal axis denotes the birth time of $1$-homological features (when holes appear) and the vertical axis represents the death time of $1$-homological features (when holes are filled). At time $\alpha = 0$, there are no $1$-simplices and consequently no $1$-holes. The first 1-cycle appears at time $\alpha=0.5$ when the 1-chain  $[a,b] + [b,c] + [c,d] + [d,a]$ appears. We mark this event by the addition of a vertical line in the persistence diagram. A second homological feature appears in the form of the cycle $[a,b] + [b,d] + [d,a]$ at time $\alpha=0.7$. This event is marked by the addition of a second vertical line in the persistence diagram. This homological feature dies at time $\alpha=0.8$ when the 2-simplex $[a,b,d]$ appears and makes the 1-cycle $[a,b] + [b,d] + [d,a]$ a 1-boundary and therefore no longer a homological feature. This is marked by the addition of the {\it point} $\bbq_1 = (0.7, 0.8)$ to the persistence diagram. At this time all simplices have been added to the filtration. This means that the homological feature $[a,b] + [b,c] + [c,d] + [d,a]$ never becomes a boundary of a 2-simplex. This is marked by adding the point $\bbq_2=(0.5, 1)$ to the persistence diagram -- recall that dissimilarities must be between $0$ and $1$ by definition.

A few more examples are shown in Figure \ref{fig_persistent_homology}. Network (a) contains one feature that appears at $\alpha=0.5$ and stays alive until $\alpha=1$. Network (b) also has one feature born at $\alpha=0.5$ that never gets trivialized until $\alpha=1$. Network (c) involves two features, born at $\alpha=0.5$ and $\alpha=0.7$ that also stay alive until $\alpha=1$. Network (d) is the same as the filtration in Figure \ref{fig_filtration}-(top). 

We close this section by noting all dissimilarity values between tuples with non-repeating elements of any dissimilarity network appear in the homological features of the induced filtration.

%
\begin{proposition}\label{prop_persistence_all}
Given a dissimilarity network $D_X^K$, any of its $k$-order dissimilarities between tuples with unique elements appear either in the death time of the $(k-1)$-th dimensional homological features or the birth time of the $k$-th dimensional features.\end{proposition}

%
\begin{myproof} See Appendix \ref{apx_proof_info}. \end{myproof}

%
Proposition \ref{prop_persistence_all} is specific to filtrations induced from dissimilarity networks since it follows from the fact that adding elements to a tuple results in {\it strictly increasing} dissimilarities [cf. Definition \ref{dfn_dissimilarity_network}]. The proposition implies that different networks result in different persistence diagrams except in the rare cases when a single point in the diagram represents multiple homological features. Thus, persistence diagrams retain almost all of the dissimilarity values of a given network and are therefore not unreasonable proxies for network discrimination. We cement this intuitive observation in the next section by proving that differences between persistence diagrams yield lower bounds on the network distances defined in Section \ref{sec_high_order_network}. We do so after an important remark.

%
\begin{remark}\label{remark_infinite}\normalfont
The maximum death time of any homological features in filtrations induced from dissimilarity networks is $1$. This is specific for filtrations constructed from dissimilarity networks because any undefined hyperedges in dissimilarity networks all take value $1$ implicitly. In computing the persistence diagram, we can either (i) set the appearance time of all undefined hyperedges as $1$ before evaluating the persistence diagrams, or (ii) determine the persistence diagram with no additional care of undefined edges, and then set the death time of any undead features to $1$. These two procedures yield the same persistence diagrams. We use the latter method in practice because it is computational simpler.
\end{remark}

%
\section{Persistence Bounds on Network Distances}\label{sec_persistence_lower_bound}
In this section we use differences between persistence diagrams to compute lower bounds of network distances. Recall that a persistence diagram $\PL k(D_X^K)$ is a collection of points of the form $\bbq = [q_b, q_d]$ where $q_b$ and $q_d>q_b$ represent the birth and death time of a $k$-homological feature [cf. Figure \ref{fig_persistent_homology}]. To compare persistence diagrams $\PL k(D_X^K)$ and $\ccalP_k\tilde\ccalL(D_Y^K)$ of networks $D_X^K$ and $D_Y^K$ we begin by defining the cost of matching features $\bbq\in\ccalQ$ and $\tbq\in\tcQ$ through the infinity norm of their difference, 
\begin{equation}\label{eqn_infty_norm}
   \|\bbq - \tbq\|_\infty := \max \big[ |q_b - \tdq_b|, |q_d - \tdq_d|\big].
\end{equation}
Further observe that the diagonal of a persistence diagram can be construed to represent an uncountable number of features with equal birth and death times. Thus, any feature $\bbq \in \ccalQ$ can be compared to any of the artificial features of the form $\tbq = [\tdq,\tdq]$. Since this feature $\tbq$ can be placed anywhere in the diagonal of the persistence diagram, we choose to place it in the point that makes the infinity norm difference smallest. This point is $\tbq = [(q_b + q_d)/2, (q_b + q_d)/2]$ which yields the difference $\|\bbq - \tbq\|_\infty = |q_d - q_b|/2$. Likewise, artificial features $\bbq = [(\tdq_b + \tdq_d)/2, (\tdq_b + \tdq_d)/2]$ can be added for any point $\tbq\in\tcQ$. We can then rephrase the comparison in \eqref{eqn_infty_norm} so that if it is more advantageous to compare with artificial diagonal features. This is formally accomplished by defining the matching cost $c(\bbq,\tbq)$ between $\bbq$ and $\tbq$ as
\begin{align}\label{eqn_cost_in_linear_bottleneck_assignment}
    c(\bbq,\tbq) \!\!:= \!\! \min\!\Big[ \! \left\|\bbq - \tbq \right\|_\infty \!, 
                    (1/2)\!\max\!\big[ \left|   q_d -    q_b \right|\!, 
                                   \left|\tdq_d - \tdq_b \right| \! \big]\!  \Big],
\end{align}
The norm $\left\|\bbq - \tbq \right\|_\infty$ is the cost of directly matching features $\bbq$ and $\tbq$. The term $(1/2)\max\big[ \left|   q_d - q_b \right|, \left|\tdq_d - \tdq_b \right|$ is the cost of matching both, $\bbq$ and $\tbq$ to artificial diagonal features. The cost $c(\bbq,\tbq)$ of matching $\bbq$ and $\tbq$ is the smaller of these two. 

If the respective collections of features $\ccalQ$ and $\tcQ$ contain the same number of elements $m=\tdm$ we can consider bijections $\pi: \ccalQ \rightarrow \tcQ$ from $\ccalQ$ to $\tcQ$. The bottleneck distance between the persistence diagrams $\PL k(D_X^K)$ and $\ccalP_k\tilde\ccalL(D_Y^K)$ of networks $D_X^K$ and $D_Y^K$ is then defined as
\begin{align}\label{eqn_bottleneck_distance_equal_elements}
    b^k(D_X^K, D_Y^K) := \min_{\pi: \ccalQ \rightarrow \tcQ} \max_{\bbq \in \ccalQ} c(\bbq, \pi(\bbq)).
\end{align}
The distance in \eqref{eqn_bottleneck_distance_equal_elements} is set to the pair of features $\bbq$ and $\tbq$ that are most difficult to match across all possible bijections $\pi$.

When the number of points in the persistence diagrams are different we assume without loss of generality that $\tdm<m$. In this case we extend $\tcQ$ by adding $\tdm-m$ diagonal features to define the set $\tcQ_e:=\tcQ \cup \{(\tdq_i,\tdq_i)\}_{i=1}^{\tdm-m}$. Since the sets $\ccalQ$ and $\tcQ_e$ contain the same number of elements bijections  $\pi: \ccalQ \rightarrow \tcQ$ are well defined. We can then modify \eqref{eqn_bottleneck_distance_equal_elements} to define a valid comparison between $\PL k$ and $\ccalP_k\tilde\ccalL$ when the number of points in the diagrams are possibly different. We do so in the following formal definition.

%
\begin{definition}\label{dfn_bottleneck}
Given persistence diagrams $\PL k(D_X^K)$ and $\ccalP_k\tilde\ccalL(D_Y^K)$ with sets of points $\ccalQ$ and $\tcQ$ having cardinalities $\tdm<m$, define the extended set $\tcQ_e:=\tcQ \cup \{(\tdq_i,\tdq_i)\}_{i=1}^{\tdm-m}$ by adding $\tdm-m$ artificial diagonal features. The bottleneck distance between the persistence diagrams $\PL k(D_X^K)$ and $\ccalP_k\tilde\ccalL(D_Y^K)$ of networks $D_X^K$ and $D_Y^K$ is defined as
\begin{align}\label{eqn_bottleneck_distance}
    b^k (D_X^K, D_Y^K) := \min_{\pi: \ccalQ \rightarrow \tcQ_e}  \max_{\bbq \in \ccalQ} c(\bbq, \pi(\bbq)),
\end{align}
where $\pi$ ranges over all bijections from $\ccalQ$ to $\tcQ_e$ and the cost $c(\bbq, \pi(\bbq))$ is defined in \eqref{eqn_cost_in_linear_bottleneck_assignment}.
\end{definition}

%
The number of bijections between two sets of points is factorial and it appears that the problem as in \eqref{eqn_bottleneck_distance} is as difficult as the problem of finding the correspondence in evaluating the network distance. However, the problem in \eqref{eqn_bottleneck_distance} is a instantiation of the Linear Bottleneck Assignment Problem (LBAP) that can be solved efficiently -- see Section \ref{sec_numerical} and \cite[Algorithm 6.1]{Burkard09}. We emphasize that $c(\bbq, \tbq) = c(\bbq, \tbq')$ for any $\bbq$ whenever $\tbq$ and $\tbq'$ are on the diagonal. Therefore, the locations of diagonal points added to construct $\ccalQ_e$ are unsubstantial as per their cost.

We prove now that the bottleneck distance between the persistence diagrams of the filtrations induced by two dissimilarity networks is a lower bound of their dissimilarity network distance.

%
\begin{theorem} \label{thm_DN}
Let $D_X^K$ and $D_Y^K$ be two $K$-order dissimilarity networks. The bottleneck distance between the $k$-th dimensional persistence diagrams of the filtrations $\ccalL( D_X^K)$ and $\ccalL( D_Y^K)$ is at most $d_{\ccalD, \infty} (D_X^K, D_Y^K)$ for any $0 \le k \le K$, i.e.
\begin{align}\label{eqn_thm_DN}
    b^k (D_X^K, D_Y^K) \leq d_{\ccalD, \infty}(D_X^K, D_Y^K).
\end{align}
\end{theorem}

%
\begin{myproof}
In order to prove Theorem \ref{thm_DN}, we introduce the following notion of augmented networks. This definition solves the issue when a single node in one network has multiple correspondents in the other network. In such cases, say, if $a$ in $X$, has both $c$ and $d$ in $Y$ as correspondents, we examine the difference between $r_X^1(a,a)$ and $r_Y^1(c,d)$ in evaluating the network difference. However $r_X^1(a,a) = r_X^0(a)$ represents the birth time of $[a]$ while $r_Y^1(c,d)$ denotes the time of appearance of an edge $[c,d]$. The augmented networks are introduced to resolve this discrepancy.

%
\begin{definition}\label{dfn_augmented_network}
Given two $K$-order dissimilarity networks $D_X^K$ and $D_Y^K$ and a correspondence $C$ between their node sets $X$ and $Y$, the augmented networks $A^K_{X, C}$ and $A^K_{Y, C}$ are a pair of $K$-order networks defined on $C$. Each node $a_i$ in $A_{X, C}^K$ and $A_{Y, C}^K$ represents a correspondent pair $(x_{C_i}, y_{C_i})$ in $C$. Relationship functions for $a_{0:k}$ with $0 \le k \le K$ are defined as
\begin{align}\label{eqn_augmented_network}
   r_{A_X}^k(a_{0:k}) = r_X^k(x_{C_0:C_k}), \ \ r_{A_Y}^k(a_{0:k}) = r_Y^k(y_{C_0:C_k}).
\end{align}\end{definition}

%
%
\begin{figure}[t]
\centering
\centerline{\def \thisplotscale {0.45}
\def \unit {\thisplotscale cm}

\begin{tikzpicture}[-stealth, shorten >=0, x = 1*\unit, y=0.5*\unit, font=\scriptsize]

    
    \node at (-2,2) { $D_X^1$: };
    \node [blue vertex, minimum height = (1.2 - 0.3)*\unit, minimum width  = (1.2 - 0.3)*\unit] at (2, 4) (x1) {$x_1$};
    \node at (1, 4) {$0.3$};
    \node [blue vertex, minimum height = (1.2 - 0.1)*\unit, minimum width  = (1.2 - 0.1)*\unit] at (4,  0) (x2) {$x_2$};    
    \node at (4,-2) {$0.1$};
    \node [blue vertex, minimum height = (1.2 - 0.1)*\unit, minimum width  = (1.2 - 0.1)*\unit] at (0,  0) (x3) {$x_3$}; 
    \node at (0,-2) {$0.1$};

    \path [-, right, line width = 0.3pt] (x1) edge node {{$0.9$}} (x2);	
    \path [-, below, line width = 2.4pt] (x2) edge node {{$0.2$}} (x3);	
    \path [-, left, line width = 0.6pt] (x3) edge node {{$0.8$}} (x1);

    \node at (8,2) { $D_Y^1$: };
    \node [blue vertex, minimum height = (1.2 - 0.3)*\unit, minimum width  = (1.2 - 0.3)*\unit] at (12, 4) (y1) {$y_1$};
    \node at (13, 4) {$0.3$};
    \node [blue vertex, minimum height = (1.2 - 0.1)*\unit, minimum width  = (1.2 - 0.1)*\unit] at (10,  0) (y3) {$y_3$}; 
    \node at (10,-2) {$0.1$};
    
    \path [-, left, line width = 0.9pt] (y3) edge node {{$0.7$}} (y1);		
    
     
       \draw (x1) edge [<->, bend left=10, below, red, thick, pos=0.6] node {$C$} (y1);
       \draw (x2) edge [<->, bend right=15, below, red, thick, pos=0.45] node {$C$} (y3);
       \draw (x3) edge [<->, bend left=13.5, above, red, thick, pos=0.6] node {$C$} (y3);

    
    \node at (-2,-5.5) { $A_X^1$: };
    \node [blue vertex, minimum height = (1.2 - 0.3)*\unit, minimum width  = (1.2 - 0.3)*\unit] at (2, -3.5) (zx1) {$a_1$};
    \node at (1,-3.5) {$0.3$};
    \node [blue vertex, minimum height = (1.2 - 0.1)*\unit, minimum width  = (1.2 - 0.1)*\unit] at (4,  -7.5) (zx2) {$a_2$};    
    \node at (5,-7.5) {$0.1$};
    \node [blue vertex, minimum height = (1.2 - 0.1)*\unit, minimum width  = (1.2 - 0.1)*\unit] at (0,  -7.5) (zx3) {$a_3$}; 
    \node at (-1,-7.5) {$0.1$};

    \path [-, right, line width = 0.3pt] (zx1) edge node {{$0.9$}} (zx2);	
    \path [-, below, line width = 2.4pt] (zx2) edge node {{$0.2$}} (zx3);	
    \path [-, left, line width = 0.6pt] (zx3) edge node {{$0.8$}} (zx1);

    
    \node at (8,-5.5) { $A_Y^1$: };
    \node [blue vertex, minimum height = (1.2 - 0.3)*\unit, minimum width  = (1.2 - 0.3)*\unit] at (12, -3.5) (zy1) {$a_1$};
    \node at (11,-3.5) {$0.3$};
    \node [blue vertex, minimum height = (1.2 - 0.1)*\unit, minimum width  = (1.2 - 0.1)*\unit] at (14,  -7.5) (zy2) {$a_2$};    
    \node at (15,-7.5) {$0.1$};
    \node [blue vertex, minimum height = (1.2 - 0.1)*\unit, minimum width  = (1.2 - 0.1)*\unit] at (10,  -7.5) (zy3) {$a_3$}; 
    \node at (9,-7.5) {$0.1$};

    \path [-, right, line width = 0.9pt] (zy1) edge node {{$0.7$}} (zy2);	
    \path [-, below, line width = 2.7pt] (zy2) edge node {{$0.1$}} (zy3);	
    \path [-, left, line width = 0.9pt] (zy3) edge node {{$0.7$}} (zy1);

\end{tikzpicture} \vspace{-3mm}}
\caption{An example to construct augmented networks. $D_X^1$ and $D_Y^1$ are $1$-order dissimilarity networks. The augmented network using correspondence $C$ induces $A_X^1 \cong D_X^1$. An additional node $a_2$ and two edges are augmented in $A_Y^1$.\vspace{-3mm}
}
\label{fig_augmented_network}
\end{figure}
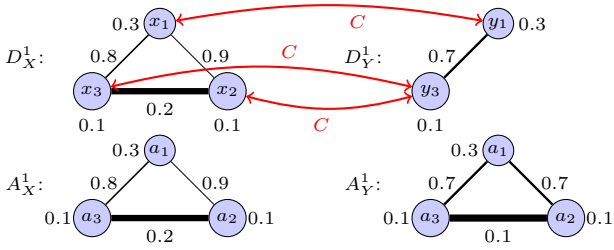

%
%
When the underlying correspondence is clear, $C$ is omitted in the subscripts. For a pair of dissimilarity networks $D_X^K$ and $D_Y^K$ and a correspondence $C$ between $X$ and $Y$, the augmented networks $A_X^K$ and $A_Y^K$ have identical $|C|$ nodes where $|C|$ denotes the number of correspondence pairs in $C$. Each node $a_i$ in both $A_X^K$ and $A_Y^K$ represents a correspondent pair $(x_{C_i}, y_{C_i})$ in $C$. For each tuple $a_{0:k}$, its relationship $r_{A_X}^k(a_{0:k})$ for the network $A_X^K$ is the same as the relationship $r_X^k(x_{C_0:C_k})$ between the tuple $x_{C_0:C_k}$ in $D_X^K$. An example to construct augmented networks is illustrated in Figure \ref{fig_augmented_network}. Augmented networks satisfy the following properties.

%
%
\begin{fact}\label{prop_augmented_network}
Given a $K$-order dissimilarity network $D_X^K$ and a correspondence $C$ between the node sets $X$ and $Y$, the augmented network $A_X^K$ constructed by Definition \ref{dfn_augmented_network} is a valid $K$-order network. The induced filtration $\ccalL(A_X^K)$ is a valid filtration and has identical persistence diagrams as $\ccalL(D_X^K)$, i.e. for any $0 \leq k \leq K$,
\begin{align}\label{eqn_prop_augmented_network}
    \PL k(A_X^K) = \PL k(D_X^K).
\end{align}\end{fact}

%
%
\begin{myproof}
See Appendix \ref{apx_proof_lower}.
\end{myproof}

%
%
Back to the proof of Theorem \ref{thm_DN}. Let $\delta = d_{\ccalD, \infty}(D_X^K, D_Y^K)$. From Definitions \ref{dfn_d_N_norm}, there exists a correspondence $C$ between $X$ and $Y$ such that $|r_X^k(x_{0:k}) - r_Y^k(y_{0:k})| \leq \delta$ for any $k$ and any pairs of correspondents $(x_{0:k}, y_{0:k}) = (x_0, y_0), \dots, (x_k, y_k) \in C$. Construct the pair of augmented networks $A_X^K$ and $A_Y^K$ defined in Definition \ref{dfn_augmented_network} using this correspondence. It follows from \eqref{eqn_augmented_network} that given a tuple $a_{0:k} \in C$ with each node $a_i$ representing the correspondent pair $(x_{C_i}, y_{C_i})$ in $C$, 
\begin{align}\label{eqn_proof_thm_DN_epsilon}
    \left| r_{A_X}^k(a_{0:k}) \! -\! r_{A_Y}^k(a_{0:k}) \right| \! =
         \! \left|r_X^k(x_{C_0:C_k}) \! - \! r_Y^k(y_{C_0:C_k}) \right| \! \leq \! \delta.
\end{align}
Since $A_X^K$ and $A_Y^K$ have identical nodes, \eqref{eqn_proof_thm_DN_epsilon} implies that any simplices that appear at time $\alpha$ in the induced filtration $\ccalL(A_X^K)$ will appear no earlier than $\alpha - \delta$ and no later than $\alpha + \delta$ in the induced filtration $\ccalL(A_Y^K)$. Using the interleaved theorem \cite[Prop. 4.2]{Chazal14}, this yields the bound
\begin{align}\label{eqn_proof_thm_DN_final}
    b^k(A_X^K, A_Y^K) \leq \delta = d_{\ccalD, \infty}(D_X^K, D_Y^K).
\end{align}
From Fact \ref{prop_augmented_network}, $b^k(A_X^K, A_Y^K) = b^k(D_X^K, D_Y^K)$. Substituting this into \eqref{eqn_proof_thm_DN_final} concludes the proof.\end{myproof}

%
%
When comparing high order networks via their induced persistence diagrams, Theorem \ref{thm_DN} provides justification that the dissimilarity obtained from persistent homologies is a lower bound on their infinity norm network distance. In particular, Theorem \ref{thm_DN} implies that: (i) A large difference in persistent homologies imply the networks to be highly different. (ii) Lower bounds can be used to estimate distance intervals because upper bounds are easy to determine using specific correspondences as per Definition \ref{dfn_d_N}. 

We note that the differences between persistence diagrams do capture important differences between networks. E.g., in Figure \ref{fig_persistent_homology} we can consider network (b) as a modification of network (a) in which we separate node $a$ into the closely related nodes $a$ and $b$ that have a dissimilarity $r_X^1(a,b) = 0.1$. The persistence diagrams of these two networks are identical, which is consistent with the relative proximity of these two networks. Network (c) is more different, because the addition of the dissimilarity  $r_X^1(b,d) = 0.7$ complicates the argument that nodes $a$ and $b$ can be well represented by a single node as in network (a). Network (d) can be argued to be in between networks (b) and (c) since the dissimilarity $r_X^1(a,b,d) = 0.8$ implies a sense of added proximity between nodes $a$, $b$, and $d$. The difference between the persistence diagrams of networks (c) and (b) is indeed larger than the difference between the diagrams of networks (d) and (b).

%
\begin{figure}[t]
\centering
\centerline{\def \thisplotscale {0.45}
\def \unit {\thisplotscale cm}

\begin{tikzpicture}[-stealth, shorten >=0, x = 1*\unit, y=0.4*\unit, font=\scriptsize]

    
    \node at (-2,2) { $D_X^1$: };
    \node [blue vertex, minimum height = (1.2 - 0)*\unit, minimum width  = (1.2 - 0)*\unit] at (2, 4) (x1) {$x_1$};
    \node at (1, 4) {$0$};
    \node [blue vertex, minimum height = (1.2 - 0.2)*\unit, minimum width  = (1.2 - 0.2)*\unit] at (4,  0) (x2) {$x_2$};    
    \node at (4,-2) {$0.2$};
    \node [blue vertex, minimum height = (1.2 - 0.12)*\unit, minimum width  = (1.2 - 0.12)*\unit] at (0,  0) (x3) {$x_3$}; 
    \node at (0,-2) {$0.12$};

    \path [-, right, line width = (1 - 0.32) * 3 pt] (x1) edge node {{$0.32$}} (x2);	
    \path [-, below, line width = (1 - 0.6) * 3 pt] (x2) edge node {{$0.6$}} (x3);	
    \path [-, left, line width = (1 - 0.42) * 3 pt] (x3) edge node {{$0.42$}} (x1);

    \node at (8,2) { $D_Y^1$: };
    \node [blue vertex, minimum height = (1.2 - 0.1)*\unit, minimum width  = (1.2 - 0.1)*\unit] at (12, 4) (y1) {$y_1$};
    \node at (13, 4) {$0.1$};
    \node [blue vertex, minimum height = (1.2 - 0.25)*\unit, minimum width  = (1.2 - 0.25)*\unit] at (10,  0) (y3) {$y_3$}; 
    \node at (10,-2) {$0.25$};
    \node [blue vertex, minimum height = (1.2 - 0.21)*\unit, minimum width  = (1.2 - 0.21)*\unit] at (14,  0) (y2) {$y_2$}; 
    \node at (14,-2) {$0.21$};
    
    \path [-, left, line width = (1 - 0.39) * 3 pt] (y3) edge node {{$0.39$}} (y1);	
    \path [-, below, line width = (1 - 0.5) * 3 pt] (y3) edge node {{$0.5$}} (y2);	
    \path [-, right, line width = (1 - 0.51) * 3 pt] (y2) edge node {{$0.51$}} (y1);		
    
     
       \draw (x1) edge [<->, bend left=10, below, red, thick, pos=0.6] node {$C$} (y1);
       \draw (x2) edge [<->, bend left=15, below, red, thick, pos=0.3] node {$C$} (y3);
       \draw (x3) edge [<->, bend right=12.5, above, red, thick, pos=0.6] node {$C$} (y2);

\end{tikzpicture} \vspace{-3mm}}
\caption{An example where the bottleneck distance between the $k$-th dimensional persistence diagrams of the filtrations $\ccalL(D_X^K)$ and $\ccalL(D_Y^K)$ is the same as their $\infty$-norm network distance for $k \in \{0, 1\}$. The optimal correspondence $C$ yields $d_{\ccalD, \infty}(D_X^1, D_Y^1) = 0.1$. The points in the $0$-dimension persistence diagrams for $\ccalL(D_X^1)$ are with coordinates $(0, \infty), (0.12, 0.42), (0.2, 0.32)$ and for $\ccalL(D_Y^1)$ are $(0.1, \infty), (0.21, 0.51), (0.25, 0.39)$. The point in the $1$-dimension persistence diagrams for $\ccalL(D_X^1)$ is with coordinate $(0.6, \infty)$ and for $\ccalL(D_Y^1)$ is $(0.5, \infty)$. The bottleneck distances between the $0$- as well as the $1$- dimensional persistence diagrams of the filtrations induced from the two networks are $0.1$.\vspace{-3mm}
}
\label{fig_tight_bound}
\end{figure}
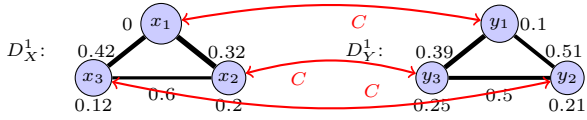
%

%
%
Further evidence for the usefulness of the bound in \eqref{eqn_thm_DN} follows from the fact that the bound is tight in some cases. An example of a tight bound is shown in Figure \ref{fig_tight_bound}. The optimal correspondence $C$, shown in the figure, yields a network distance $d_{\ccalD, \infty}(D_X^1, D_Y^1) = 0.1$. The coordinates of the points in the $0$-dimension persistence diagrams for $\ccalL(D_X^1)$ are $(0, \infty), (0.12, 0.42), (0.2, 0.32)$ and for $\ccalL(D_Y^1)$ are $(0.1, \infty), (0.21, 0.51), (0.25, 0.39)$. The coordinates of the point in the $1$-dimension persistence diagrams for $\ccalL(D_X^1)$ are $(0.6, \infty)$ and for $\ccalL(D_Y^1)$ is $(0.5, \infty)$. The bottleneck distances between the $0$- and $1$-dimensional persistence diagrams of the induced filtrations are $0.1$; same as $d_{\ccalD,\infty}(D_X^K, D_Y^K)$.

%
\subsection{Persistence Bounds on $k$-order Distances}\label{sec_persistence_lower_bound_k}
Theorem \ref{thm_DN} guarantees that the $\infty$-norm dissimilarity network distance $d_{\ccalD, \infty}$ can be tightly bounded by persistence diagrams. In this section we build on Theorem \ref{thm_DN} to show that the $k$-order network distance $d_\ccalD^k$ can also be bounded by persistent methods. 

%
\begin{theorem} \label{thm_DN_k}
Given two dissimilarity networks $D_X^K$ and $D_Y^K$ and an integer $1 \le k \le K$, the bottleneck distance between the $k'$-th dimensional persistence diagrams of the filtrations $\ccalL(D_X^K)$ and $\ccalL(D_Y^K)$ is at most $d_\ccalD^k(D_X^K, D_Y^K)$ for any $0 \leq k' < k$, i.e.
\begin{align}\label{eqn_thm_DN_k}
    b^{k'}(D_X^K, D_Y^K) \leq d_\ccalD^k(D_X^K, D_Y^K).
\end{align}
\end{theorem}

%
%
\begin{myproof} We prove Theorem \ref{thm_DN_k} from Theorem \ref{thm_DN}. To do that, we leverage the relationship for network distances as we present next.

%
%
\begin{fact} \label{prop_d_relationship}
Given high order networks $N_X^K$ and $N_Y^K$, we have
\begin{align}\label{eqn_prop_d_relationship}
    d_\ccalN^{k'}(N_X^K, N_Y^K) \leq d_\ccalN^{k}(N_X^K, N_Y^K),
\end{align}
for $0 \leq k' \leq k - 1 \leq K$, and
\begin{align}\label{eqn_prop_d_relationship_2}
    d_{\ccalN, \infty}(N_X^K, N_Y^K) = d_\ccalN^K(N_X^K, N_Y^K).
\end{align}\end{fact}

%
%
\begin{myproof}
See Appendix \ref{apx_proof_lower}.
\end{myproof}

%
Fact \ref{prop_d_relationship} implies that $d_\ccalN^k$ increases as the order $k$ becomes higher, and the $\infty$-norm distance $d_{\ccalN,\infty}$ is the same as the $K$-order network distance. The result in Theorem \ref{thm_DN_k} only considers relationship functions of order $k$, and to leverage the connection as in Fact \ref{prop_d_relationship}, we introduce the following notion of truncated networks.

%
%
\begin{definition}\label{dfn_truancated_network}
Given $K$-order network $N_X^K = (X, r_X^0, \dots, r_X^K)$, its $k$-order truncated network $N_X^k$ is defined as $N_X^k = (X, r_X^0, \dots, r_X^k)$.
\end{definition}

%
%
A $k$-order truncated network $N_X^k$ has the same node set as its parent network $N_X^K$ and collects the lowest $k+1$ order relationship functions of $N_X^K$. Back to the proof of Theorem \ref{thm_DN_k}, construct the $k$-order truncated networks $D_X^k$ and $D_Y^k$ from $D_X^K$ and $D_Y^K$. It follows directly from Theorem \ref{thm_DN} that for any $0 \leq k' \leq k - 1$, 
\begin{align}\label{eqn_proof_thm_DN_k_start}
    b^{k'}(D_X^K, D_Y^K) \leq d_{\ccalD, \infty}(D_X^k, D_Y^k).
\end{align}
Since $D_X^k$ and $D_Y^k$ are valid high order networks, Fact \ref{prop_d_relationship} implies that $d_{\ccalD, \infty}(D_X^k, D_Y^k) = d_{\ccalD}^k (D_X^k, D_Y^k)$. Meanwhile, $d_{\ccalD}^k (D_X^k, D_Y^k) = d_{\ccalD}^k (D_X^K, D_Y^K)$ follows from the facts that $D_X^k$ and $D_X^K$ have identical $k$-order relationship functions and $D_Y^k$ and $D_Y^K$ have same $k$-order relationship. Finally, for any $k' \leq k - 1$, the $k'$-th dimensional persistence diagram of $\ccalL(D_X^k)$ is identical to that of $\ccalL(D_X^K)$ and the $k'$-th dimensional persistence diagram of $\ccalL(D_Y^k)$ is identical to that of $\ccalL(D_Y^K)$, therefore $ b^{k'}(D_X^k, D_Y^k) = b^{k'}(D_X^K, D_Y^K)$. Combining these observations, the proof concludes. \end{myproof}

%
Theorem \ref{thm_DN_k} bounds the $k$-order network distance by the bottleneck distance between persistence diagrams of any order $k' < k$. We note that best lower bound is not necessarily $k'=k-1$. Observe that the result in Theorem \ref{thm_DN_k} does not apply for $k=0$. This is not a problem because the $0$-order network distances $d_\ccalD^0$ examine relationships between individual nodes only. Thus, we can compute the optimal correspondence in Definition \ref{dfn_d_N} by matching each of the relationships $r_X^0(x)$ in the network $\ccalD_X^K$ to the closest relationship $r_Y^0(y)$ in the network $\ccalD_X^K$. We emphasize that the lower bound described in Theorem \ref{thm_DN_k} is also tight since we can find dissimilarity networks $D_X^K$ and $D_Y^K$ such that $d_\ccalD^k(D_X^K, D_Y^K)$ equals the bottleneck distance between the $k'$-th dimensional persistence diagrams of the filtration $\ccalL(D_X^k)$ and $\ccalL(D_Y^k)$ for any $0 \leq k' \leq k - 1 \leq K$. See Figure \ref{fig_tight_bound} for an illustration where bottleneck distances between the $k$-dimensional persistence diagrams coincide with $d_\ccalD^k$. 

One of the key arguments in the proof of Theorem \ref{prop_d_relationship} is the fact that the infinity norm network distance $d_{\ccalN, \infty}(N_X^K, N_Y^K)$ and the $K$-order network distance $d_\ccalN^K(N_X^K, N_Y^K)$ coincide [cf. Fact \ref{prop_d_relationship}]. This equality implies that Theorem \ref{thm_DN} follows as a particular case of Theorem \ref{thm_DN_k} by setting $k=K$ in \eqref{eqn_thm_DN_k} and observing that $d_{\ccalN, \infty}(N_X^K, N_Y^K) = d_\ccalN^K(N_X^K, N_Y^K)$. Do notice that this doesn't mean that the proof of Theorem \ref{thm_DN} is redundant because Theorem \ref{thm_DN} is leveraged in the proof of Theorem \ref{thm_DN_k}. However, the result does imply that the infinity norm distance $d_{\ccalN, \infty}(N_X^K, N_Y^K)$ does not contain information beside the one that is contained in the collection of $k$-order distance $d_\ccalN^k(N_X^K, N_Y^K)$ for $k=0,\ldots,K$.

Another consequence of  Theorem \ref{thm_DN_k} is that we can bound the $p$-norm network distance $d_{\ccalD, p}(D_X^K, D_Y^K)$ [cf. Definition \ref{dfn_d_N_norm}] for arbitrary $p$. This can be done using the individual bounds for $d_\ccalD^k(D_X^K, D_Y^K)$ established in \eqref{eqn_thm_DN_k} as we show next.

%
\begin{corollary}\label{corollary_p_norm_bound}
Group the $k$-order dissimilarity network distances in the vector $\bbd_{\ccalD}^K (D_X^K, D_Y^K) :=$ $\big[d_\ccalD^0(D_X^K, D_Y^K),$ $d_\ccalD^1(D_X^K, D_Y^K),$ $\ldots,$ $d_\ccalD^K(D_X^K, D_Y^K)\big]^T$. Further define the vector $\bbb^K (D_X^K, D_Y^K) :=$ $\big[d_\ccalD^0(D_X^K, D_Y^K),$ $b^{k_1} (D_X^K, D_Y^K),$ $\ldots,$ $b^{k_K} (D_X^K, D_Y^K)\big]^T$ whose first component is the zero order distance $d_\ccalD^0(D_X^K, D_Y^K)$ and whose other components are bottleneck lower bounds with $k_l < l$. The $p$-norm network distance can be lower bounded as
\begin{align}\label{eqn_p_norm_network_distance_bound_final}
   \left\| \bbb^K (D_X^K, D_Y^K) \right\|_p 
       \! \leq\! \left\| \bbd_{\ccalD}^K (D_X^K, D_Y^K) \right\|_p 
       \! \leq\! d_{\ccalD, p}(D_X^K, D_Y^K).  
\end{align}
\end{corollary}

%
\begin{myproof}
The second inequality follows from the fact that a single correspondence is used in $d_{\ccalD,p}$ whereas an order-specific correspondence may be utilized in each $k$-order distance $d_\ccalD^k$ \cite{Huang15}. For the first inequality, $b^{k_l}(D_X^K, D_Y^K) \leq d_\ccalD^l(D_X^K, D_Y^K)$ comes from Theorem \eqref{thm_DN_k}. This implies that the vector $\bbb^K(D_X^K, D_Y^K)$ is element-wise smaller or equal to the vector $\bbd_\ccalD^K(D_X^K, D_Y^K)$.  \end{myproof}

%
Corollary \ref{corollary_p_norm_bound} shows that the $p$-norm network distance $d_{\ccalD,p}$ can be lower bounded using: (i) The exact value of the $0$-order network distance. (ii) The persistence homology lower bounds in \eqref{eqn_thm_DN_k} of Theorem \ref{thm_DN_k}. The bounds in Corollary \ref{corollary_p_norm_bound} are expected to be the most useful when they combine information gleaned from persistence diagrams of different orders.

%
\begin{remark}\label{rmk_epsilon_summands}\normalfont
The requirement of having the inequality in \eqref{eqn_dfn_order_increasing} be strict unless all the elements of $x_{0:k}$ appear in $x_{0:k-1}$ for the network to be a dissimilarity network [cf. Definition \ref{dfn_dissimilarity_network}] is not always naturally satisfied in practice. As we have seen, its satisfaction may necessitate the addition of a small but arbitrary constant to some tuple relationships [cf. Figure \ref{fig_dissimilarity_network_example}]. This is not a problem in practice because Theorems \ref{thm_DN} and \ref {thm_DN_k} still holds if this technical condition is violated. The only modification in the statements is that the respective network distances are not metrics but pseudometrics. What this means is that it is possible to construct pathological examples of networks that satisfy the order decreasing property {\it not} strictly and that have null network distance while {\it not} being isomorphic. These networks are unlikely to appear in practice and, even if they do, the bounds in Theorems \ref{thm_DN} and \ref {thm_DN_k} hold, albeit with a different interpretation.
\end{remark}

%
\section{Proximity Networks}\label{sec_proximity_network}

In dissimilarity networks relationships $r_X^k(x_{0:k})$ encode a level of dissimilarity between elements of the tuple. The relationship $r_X^k(x_{0:k})$ may, alternatively, describe similarity or proximity between elements. To account for this, we define proximity networks as those in which the relationship in \eqref{eqn_dfn_order_increasing} is flipped to
\begin{align}\label{eqn_dfn_order_decreasing}
   r_X^k(x_{0:k}) \leq r_X^{k-1}(x_{0:k-1}),
\end{align}
with the inequality equalized if and only if the element $x_k$ also appear in $x_{0:k-1}$. We refer to \eqref{eqn_dfn_order_decreasing} as the order decreasing property and denote the set of all proximity networks of order $K$ as $\ccalP^K$. When the input networks of Definition \ref{dfn_d_N} or \ref{dfn_d_N_norm} are proximity networks, we refer to the network distances as the $k$-order proximity network distance $d_\ccalP^k$ or the $p$-norm proximity network distances $d_\Pnorm$. The restrictions to proximities make the distances well-defined {\it metrics} in the respective spaces \cite{Huang15}. 

%
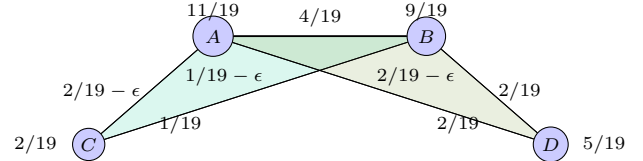
\begin{figure}[t]
\centerline{\def \thisplotscale {0.45}
\def \unit {\thisplotscale cm}

\pgfdeclarelayer{background}
\pgfdeclarelayer{foreground}
\pgfsetlayers{background,foreground}

\begin{tikzpicture}[-stealth, shorten >=0, x = 1.05*\unit, y=0.4*\unit, font=\scriptsize]

    \begin{pgfonlayer}{foreground}
       \node [blue vertex, minimum width = (0.9 + 11/38) * \unit, minimum height = (0.9 + 11/38) * \unit] at ( 2, 8) (A) {$A$}; \node at ( 2, 9.8) {$11/19 $};
       \node [blue vertex, minimum width = (0.9 + 9/38) * \unit, minimum height = (0.9 + 9/38) * \unit] at ( 8, 8) (B) {$B$}; \node at ( 8, 9.8) {$9/19 $};  
       \node [blue vertex, minimum width = (0.9 + 2/38) * \unit, minimum height = (0.9 + 2/38) * \unit] at ( -1.5, 0) (C) {$C$}; \node at (-3, 0) {$2/19 $};  
       \node [blue vertex, minimum width = (0.9 + 5/38) * \unit, minimum height = (0.9 + 5/38) * \unit] at (11.5, 0) (D) {$D$}; \node at (13, 0) {$5/19 $};  
       \path [-, above, line width = 12/19 pt] (A) edge node {{$4/19 $}} (B);	
       \path [-, left, line width = 6/19 pt] (A) edge node {{$2/19 -\epsilon$}} (C);	
       \path [-, right, pos = 0.85, line width = 3/19 pt] (B) edge node {{$1/19 $}} (C);	
       \path [-, right, line width = 6/19 pt] (B) edge node {{$2/19 $}} (D);
       \path [-, left, pos = 0.85, line width = 6/19 pt] (A) edge node {{$2/19 $}} (D);
    \end{pgfonlayer}

    \begin{pgfonlayer}{background}    
      \filldraw[ultra thin, fill = orange!30!cyan, fill opacity = 0.2]
              (A.center) --(B.center) --(C.center) -- cycle;
      \node at (2.3, 5) {$1/19 - \epsilon$};  
       \filldraw[ultra thin, fill = orange!60!cyan, fill opacity = 0.2] 
               (A.center) --(B.center) --(D.center) -- cycle;
      \node at (7.7, 5) {$2/19 - \epsilon$};  
    \end{pgfonlayer}
        
\end{tikzpicture} \vspace{-4mm}}
\caption{Collaborations between authors in a research community. The $k$-order relationship in this $2$-order network describes the number of publications between members of a $(k+1)$-tuple normalized by the total number of papers. E.g., $A$ writes $11$ papers in total, and coauthors $4$, $2$, and $2$ papers respectively with $B$, $D$, and $C$. The $\epsilon$ in the relationship function is a technical modification to differentiate relationship between two authors from that of a single author. Author $A$ also writes $1$ paper jointly with $B$ and $C$.\vspace{-4mm}
}
\label{fig_proximity_network_example}
\end{figure}
%

%
As is the case with the order increasing property in dissimilarity networks, the order decreasing  property is well justified in proximity networks. E.g., consider a different aspect of coauthorship in which the $k$-order relationship function between authors is proportional to the number of publications coauthored by all members of a given $(k+1)$-tuple. In particular, the zeroth order proximities $r_X^0$ are the numbers of papers published by authors normalized by the total number of papers. In Figure \ref{fig_proximity_network_example} authors $A, B, C, D$ publish $11$, $9$, $2$, and $5$ papers respectively and there are $19$ papers in total which implies $r_X^0(A) = 11/19$, $r_X^0(B) = 9/19$, $r_X^0(C) = 2/19$, $r_X^0(D) = 5/19$. The first order proximities $r_X^1$ represent the number of papers co-published by nodes. Since a collaboration for a pair of authors is also a paper for each of the individuals it is certain that $r_X^1(x, x') \le r_X^0(x)$. In Figure \ref{fig_proximity_network_example}, $A$ and $B$ collaborate on $4$ papers, which is less than the $11$ and $9$ papers written by each of the individuals. Authors $A$ and $C$ as well as $A$ and $D$ coauthor $2$ papers in total. Second order proximities $r_X^2$ for triplets indicate the normalized number of papers coauthored by the three members. Since a paper with three authors is also a collaboration for the three pairs of authors we must have $r_X^2(x, x', x'') \le r_X^1(x, x')$. In Figure \ref{fig_proximity_network_example}, authors $A$, $B$, and $D$ cowrite $2$ papers, which is no more than the number of pairwise collaborations between each pair of the authors. An $\epsilon$ constant is subtracted from some proximities to satisfy the technical condition that $r_X^k(x_{0:k}) = r_X^{k-1}(x_{0:k-1})$ if and only the elements in $x_{0:k}$ are all elements of $x_{0:k-1}$. This is not a concern in practice either -- see Remark \ref{rmk_epsilon_summands}. Undefined hyperedge relationships in proximity networks take the value $0$ which is the minimum possible proximity for a tuple. 

For any proximity network $P_X^K$ with relationships $p_X^k(x_{0:k})$ we can define the dual network $D_X^K$ with relationship functions $d_X^k(x_{0:k}) := 1 - p_X^k(x_{0:k})$. It is not difficult to see that $D_X^K$ is a well defined dissimilarity network and any of the network distances between the dual dissimilarity networks coincides with the respective distance between the proximity networks \cite{Huang15}. It follows that the bounds in Theorems \ref{thm_DN} and \ref{thm_DN_k} apply for proximity networks with the caveat that the filtrations $\PL k(D_X^K)$ and $\PL k(D_Y^K)$ are filtrations of the dual networks associated with the given proximity networks $P_X^K$ and $P_Y^K$.

%
\section{Implementation Details}\label{sec_numerical}

We discuss implementation issues regarding the procedure for evaluating lower bounds, its computational cost, and some heuristic simplifications.

\medskip\noindent\textit{Procedure.} To compute distance lower bounds for a dissimilarity network we construct filtrations and evaluate persistence diagrams. Algorithmic procedures to compute filtrations and persistence diagrams are discussed in \cite{Zomorodian05,davie2013}. The bottleneck distances $b^k(D_X^K, D_Y^K)$ between the persistence diagrams of different orders $0\leq k\leq K$ are then evaluated. These distances follow from Definition \ref{dfn_bottleneck} and are found as solutions of LBAP with costs given by \eqref{eqn_cost_in_linear_bottleneck_assignment}. If the interest is in the infinity norm network distances $d_{\ccalD,\infty}$, any of the bottlenecks distances is a lower bound [cf. Theorem \ref{thm_DN}]. If we are interested in the $k$-order network distance $d_{\ccalD}^k$, the bottleneck distance between persistence diagrams of dimension $k' \leq k - 1$ is a lower bound [cf. Theorem \ref{thm_DN_k}]. When given a proximity network we construct the dual network with relationships $r_X^k(x_{0:k})=1-r_X^k(x_{0:k})$ and proceed as before.

\medskip\noindent\textit{Computational cost.} The process above involves computation of the persistent homology and evaluation of the bottleneck distance. Given a $K$-order network with $n$ nodes, denote as $\tdn$ the total number of simplices and $m$ as the number of points in the persistence diagram. The complexity of computing a persistence diagram is akin to matrix multiplication on the number of simplices. This yields a computational complexity of order $O(\tdn^{2.373})$ \cite{Zomorodian05,davie2013}. The complexity of evaluating the bottleneck distance using LBAP [cf. \eqref{eqn_bottleneck_distance}] is of order $O(m^{2.5} \log m)$ \cite[Algorithm 6.1]{Burkard09}. The number of simplices of order $k$ can be as large $n^k$ and the number of points in a persistence diagram as large as the number of simplices. This means that we can have $m=\tdn = n^K$, which yields impractical computational times for orders larger than $K=2$. In the practical examples we consider in Sections \ref{sec_compare_coauthorship} and \ref{sec_engineering_different}, the number of simplices $\tdn$ is much smaller than the maximum possible and scales linearly with the number of nodes. This yields complexity $O(n^{2.373})$ for computing the persistence diagram and $O(n^{2.5} \log n)$ for the bottleneck distance.

\medskip\noindent\textit{Undefined tuple relationships.} Formally, dissimilarity networks require that all undefined tuple relationships take value $1$. As defined above this is impractical because it would scale the total number of simplices as $\tdn = n^k$. In computations we leave these simplices undefined and let features die at infinity. As it follows from Remark \ref{remark_infinite}, the death time of these features can be set to $\alpha=1$ to yield the same result that would be obtained from setting undefined tuple relationships to 1 before computing the persistence diagram.

\medskip\noindent\textit{Elimination of small homological features.} We can reduce the number of points in a persistence diagram by removing points $\bbq$ close to the diagonal. This is justified because: (i) These points represent ephemeral homological features and are likely to be generated by noise in observations. (ii) They contribute a small value to the cost in \eqref{eqn_cost_in_linear_bottleneck_assignment} and won't improve the lower bound much from the trivial nonnegative lower bounds $ 0 \leq d_{\ccalD, \infty}(D_X^K, D_Y^K)$ or $0\leq d_\ccalD^k(D_X^K, D_Y^K)$ that hold for any distance.

%
\begin{figure*}[t] 

\begin{minipage}[h]{0.24\textwidth}
    	\centering
    	\includegraphics[trim=1.2cm 0.8cm 1.2cm 0.8cm, clip=true, width=1 \textwidth]{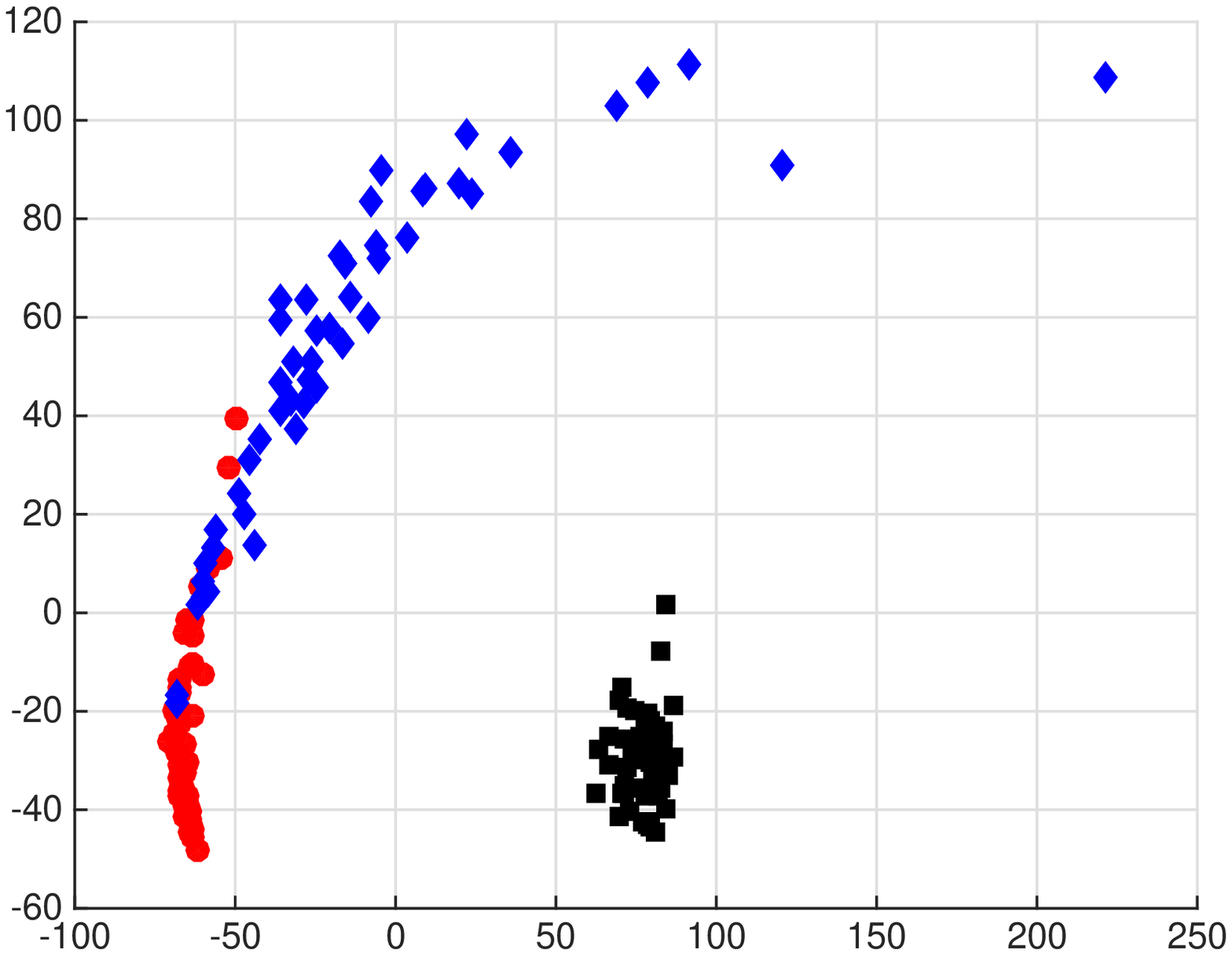}
	\scriptsize (a) $b^0$, $30$ nodes
	\end{minipage}
\begin{minipage}[h]{0.24\textwidth}
    	\centering
    	\includegraphics[trim=1.2cm 0.8cm 1.2cm 0.8cm, clip=true, width=1 \textwidth]{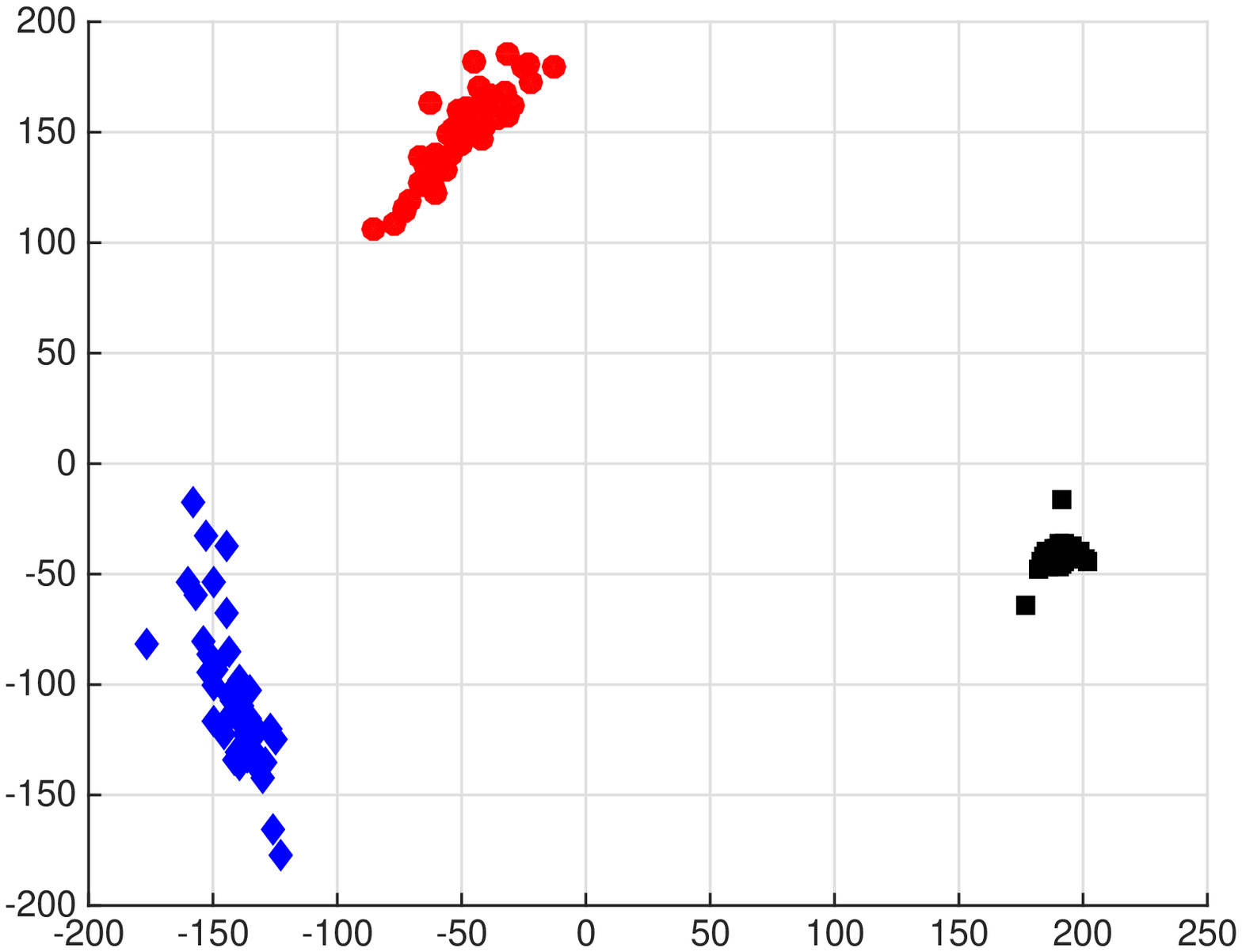}
	\scriptsize (b) $b^1$, $30$ nodes
	\end{minipage}
\begin{minipage}[h]{0.24\textwidth}
    	\centering
	\includegraphics[trim=1.2cm 0.8cm 1.2cm 0.8cm, clip=true, width=1 \textwidth]{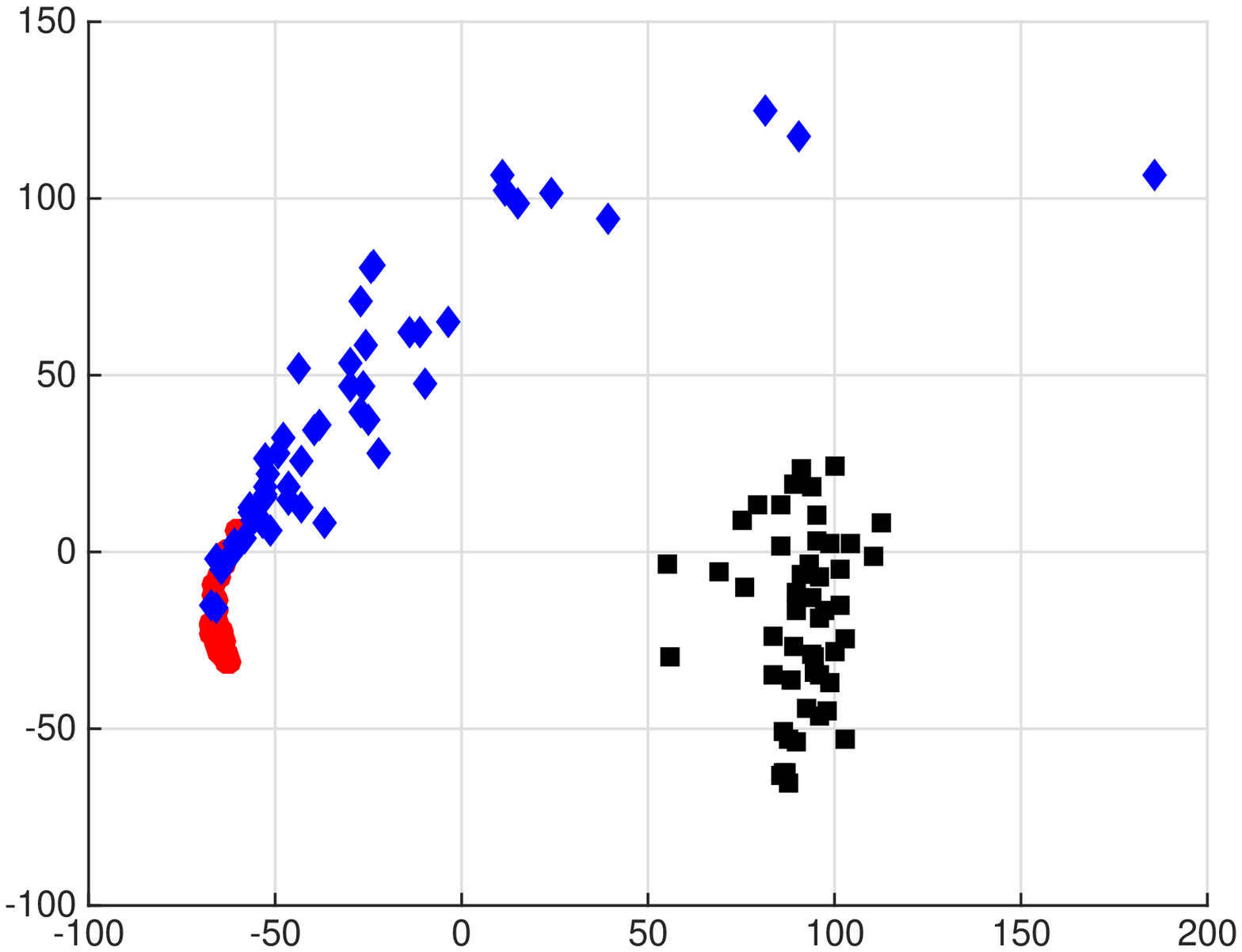}
	\scriptsize (c) $b^0$, $30$ to $39$ nodes
\end{minipage}
\begin{minipage}[h]{0.24\textwidth}
    	\centering
	\includegraphics[trim=1.2cm 0.8cm 1.2cm 0.8cm, clip=true, width=1 \textwidth]{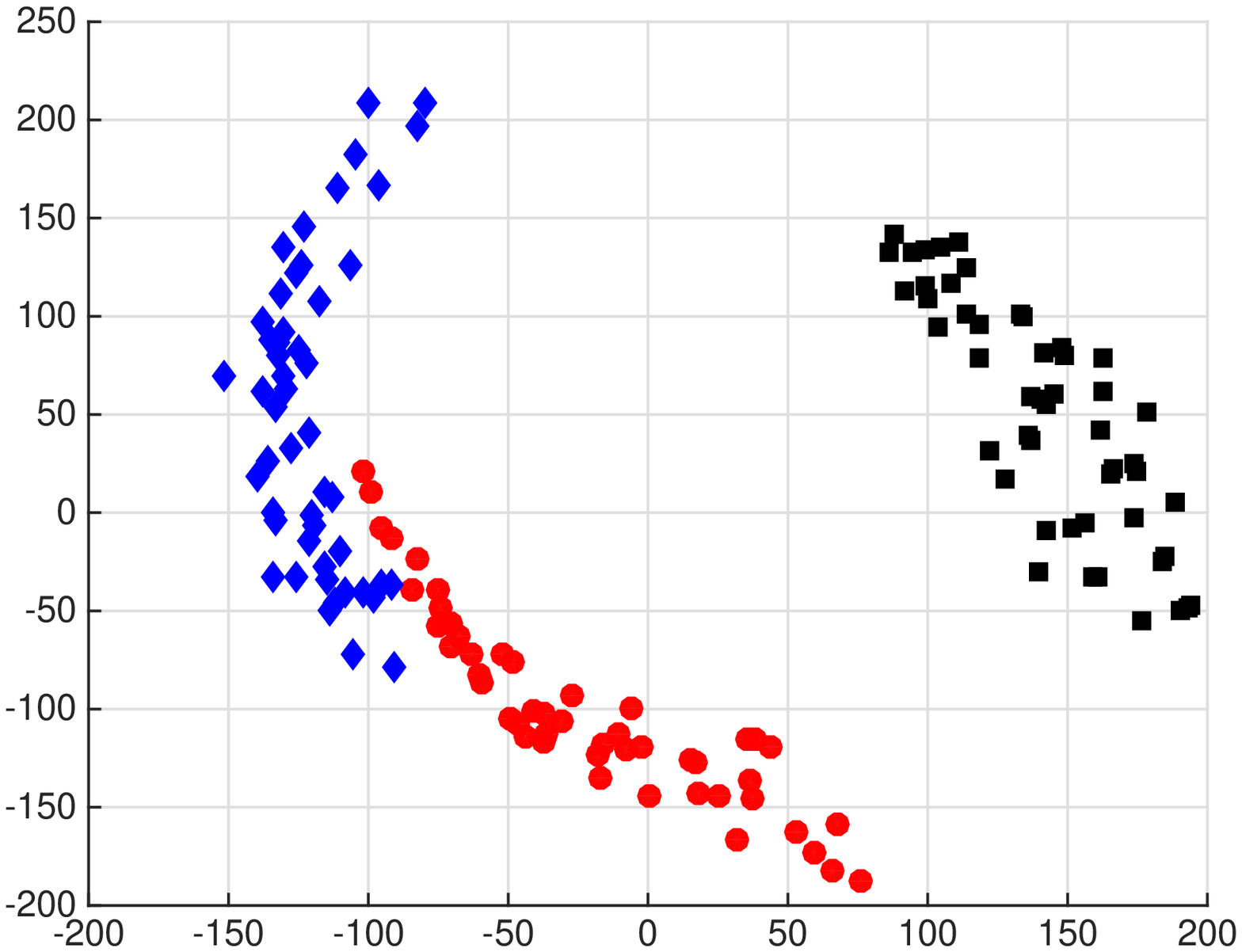}
	\scriptsize (d) $b^1$, $30$ to $39$ nodes
\end{minipage}
\caption{Two dimensional Euclidean embeddings of the networks constructed from three different models with different number of nodes with respect to the network metric lower bounds $b^0$ and $b^1$. In the embeddings, red circles denote networks constructed from the Erd\H os-R\' enyi model, blue diamonds represent networks constructed from the unit circle model, and black squares the networks from the correlation model.\vspace{-4mm}
}
\label{fig_synthetic}
\end{figure*}

%
\begin{figure*}[t] 
\begin{minipage}[h]{0.24\textwidth}
    	\centering
    	\includegraphics[trim=1.5cm 0.8cm 1.5cm 0.8cm, clip=true, width=1 \textwidth]{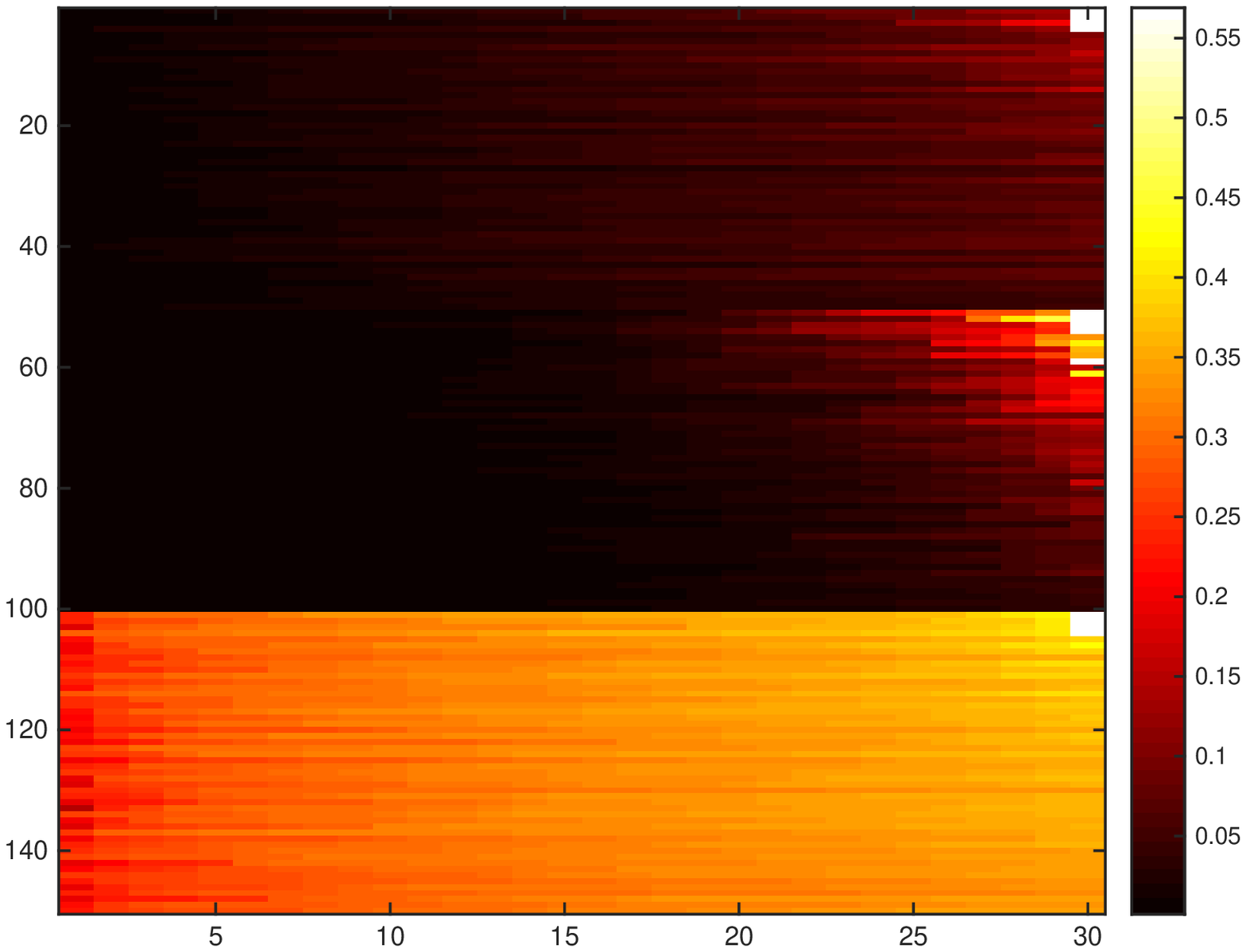}
	\scriptsize (a) death time in $\PL 0$, ascending order
	\end{minipage}
\begin{minipage}[h]{0.24\textwidth}
    	\centering
	\includegraphics[trim=1.5cm 0.8cm 1.5cm 0.8cm, clip=true, width=1 \textwidth]{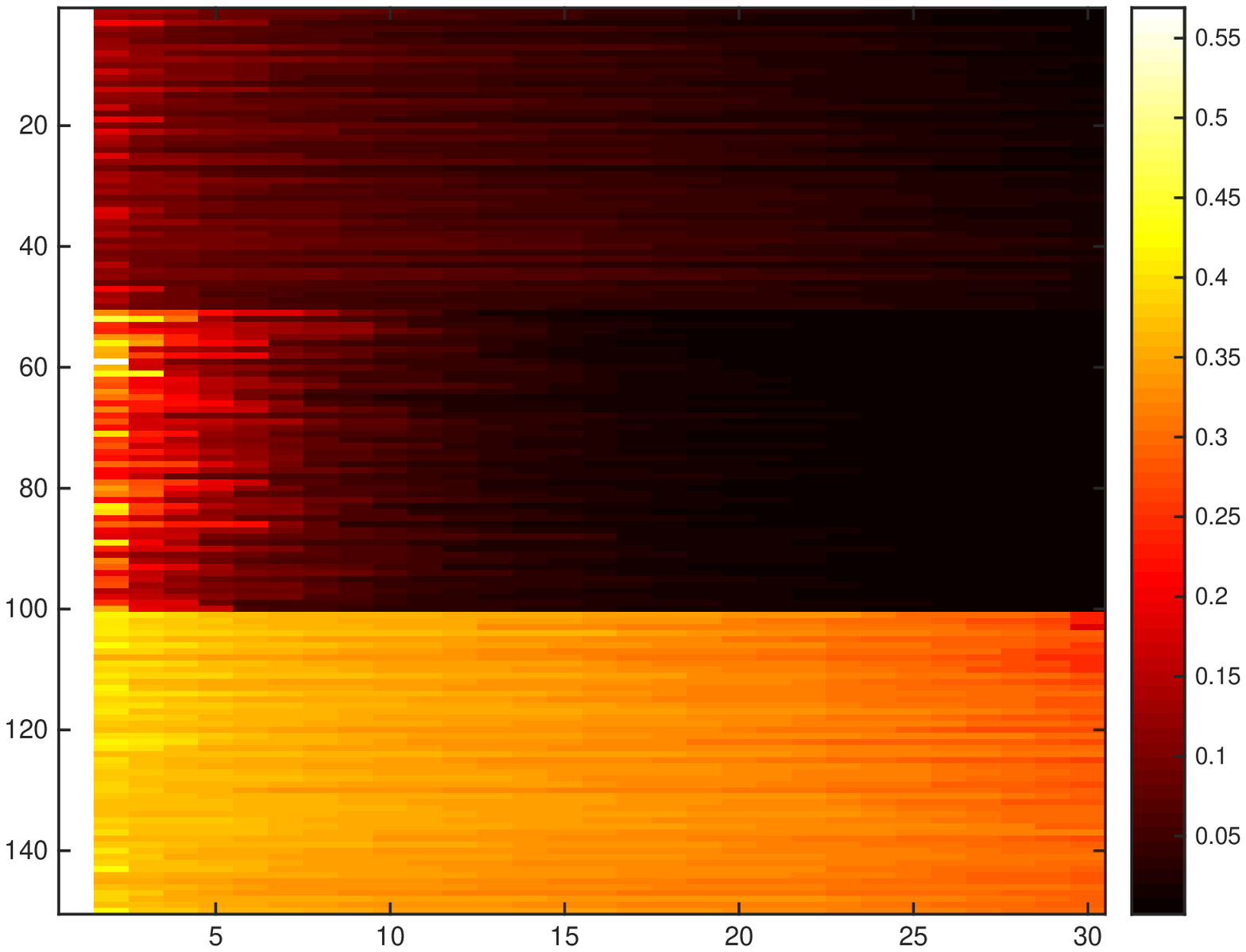}
	\scriptsize (b) death time in $\PL 0$, descending order
	\end{minipage}
\begin{minipage}[h]{0.24\textwidth}
    	\centering
	\includegraphics[trim=1.5cm 0.8cm 1.5cm 0.8cm, clip=true, width=1 \textwidth]{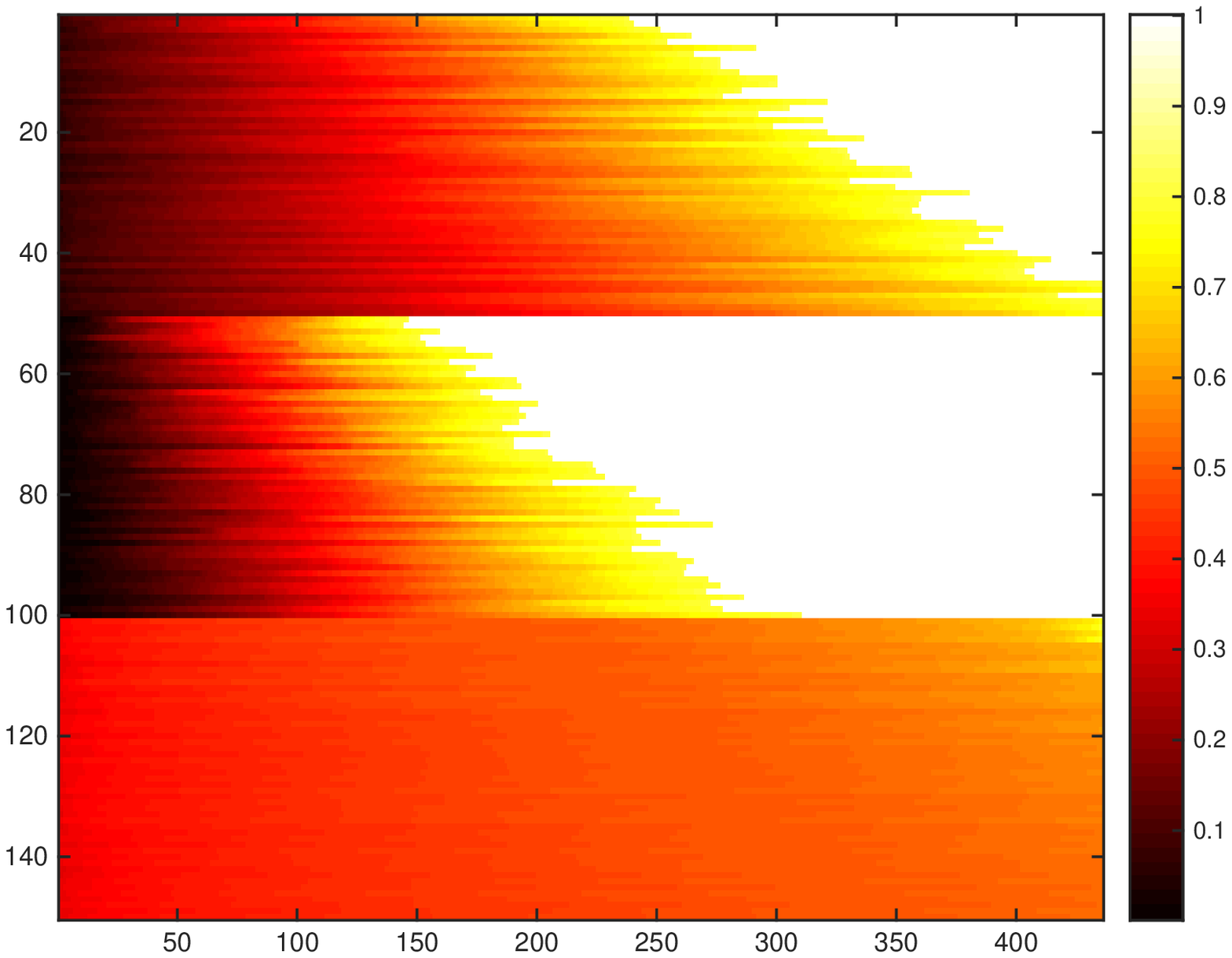}
	\scriptsize (c) birth time in $\PL 1$, ascending order 
\end{minipage}
\begin{minipage}[h]{0.24\textwidth}
    	\centering
	\includegraphics[trim=1.5cm 0.8cm 1.5cm 0.8cm, clip=true, width=1 \textwidth]{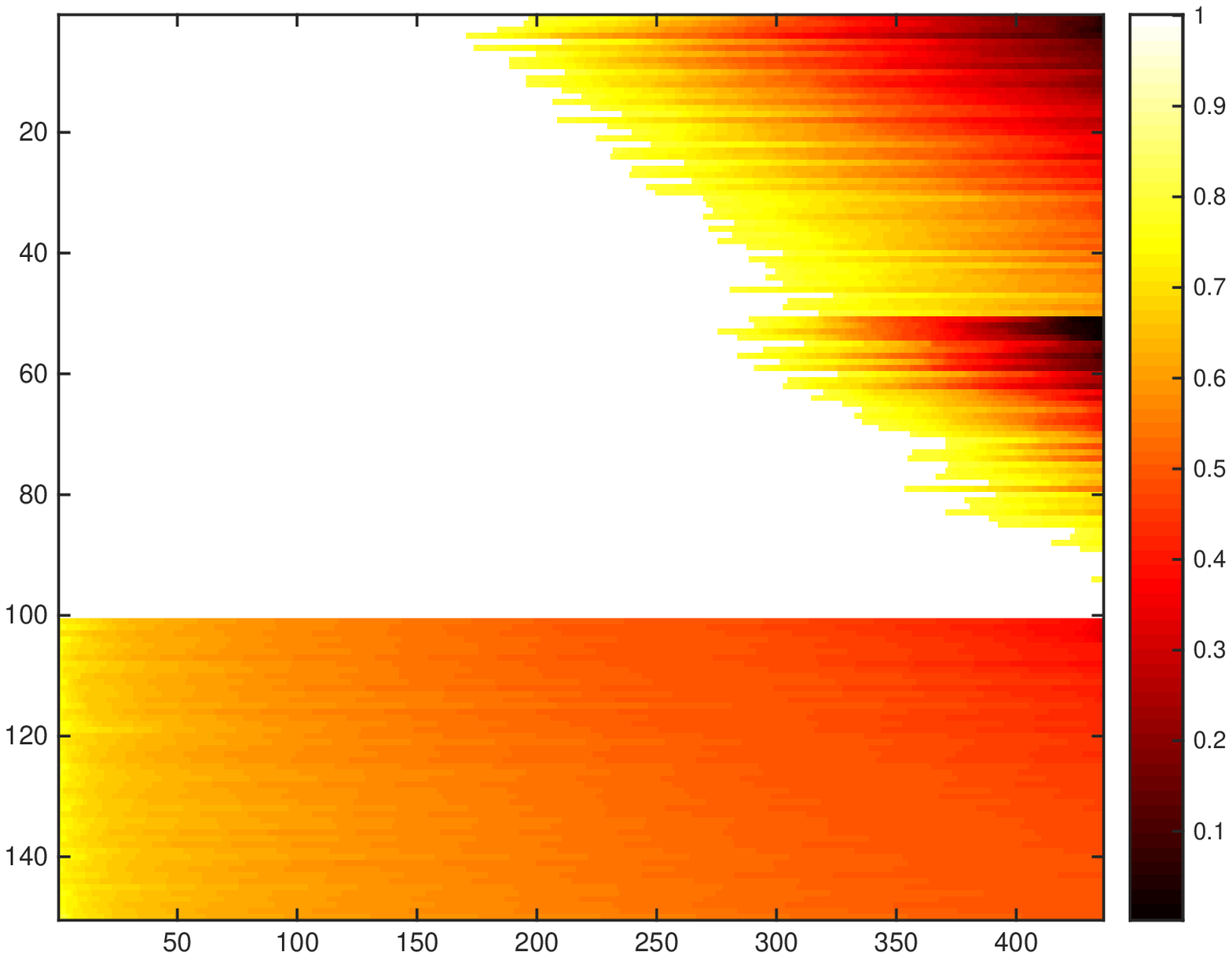}
	\scriptsize (d) birth time in $\PL 1$, descending order 
	\end{minipage}
\caption{Illustrations of the set of death time in $0$-dimensional persistence diagrams $\PL 0$ and the set of birth time in $1$-dimensional persistence diagrams $\PL 1$ for the filtrations constructed from the randomly generated networks with number of nodes ranging between $30$ and $39$. The top 50 networks represent those constructed from the Erd\H os-R\' enyi model, the middle 50 networks are with the unit circle model, and the bottom 50 are with the correlation model.\vspace{-3mm}
}
\label{fig_interpretation}
\end{figure*}

%
%
\section{Applications}\label{sec_applications}

We illustrate the usefulness of homology methods through experiments in both synthetic (Section \ref{sec_synthetic}) and real-world data (Sections \ref{sec_compare_coauthorship} and \ref{sec_engineering_different}). The objective is to demonstrate that networks with similar structures and should be alike to each other are indeed similar in their persistent homologies.

%
\subsection{Classification of Synthetic Networks}\label{sec_synthetic}

We consider three types of synthetic weighted pairwise networks. Edge weights in all three types of networks encode proximities. The first type of networks are with weighted Erd\H os-R\' enyi model \cite{erd6s1960}, where the edge weight between any pair of nodes is a random number uniformly selected from the unit interval $[0, 1]$. In the second type of networks, the coordinates of the vertices are generated uniformly and randomly in the unit square, and the edge weights are evaluated with the Gaussian radial basis function $\exp(-d(i,j)^2 / 2\sigma^2)$ where $d(i,j)$ is the distance between vertices $i$ and $j$ in the unit circle and and $\sigma$ is a kernel width parameter. In all simulations, we set $\sigma$ to $0.5$. The edge weight measures the proximity between the pair of vertices and takes value in the unit interval. In the third type of networks, we consider that each vertex $i$ represents an underlying feature $\bbu_i \in \reals^d$ of dimension $d$, and examine the Pearson's linear correlation coefficient $\rho_{ij}$ between the corresponding features $\bbu_i$ and $\bbu_j$ for a given pair of nodes $i$ and $j$. The weight for the edge connecting the pair is then set as $\rho_{ij} / 2 + 0.5$, a proximity measure in the unit interval. The feature space dimension $d$ is set as $30$ in all simulations.

We start with networks of equal size $|X| = 30$ and construct $50$ random networks for each aforementioned type. The edges with weights no greater than a threshold $\tau$ are removed to create sparsity. We set $\tau = 0.2$ in all simulations. In order to transform the constructed pairwise networks into high order proximity networks, the $0$-order proximity for any node is set to $0$, i.e. $r_X^0(x) := 0$ for any $x \in X$. We then use the persistent homology method described in Section \ref{sec_numerical} to evaluate the dissimilarities between all networks. The persistent homologies are computed using JavaPlex \cite{JPlex}. Figure \ref{fig_synthetic} (a) and (b) plot the two dimensional Euclidean embeddings \cite{CoxCox08} of the network metric lower bounds $b^0$ and $b^1$ between the $0$- and $1$-dimensional persistence diagrams respectively. Networks constructed with different models form clear separate clusters with respect to $b^1$ where networks with Erd\H os-R\' enyi model are denoted by red circles, networks with unit circle model are described by blue diamonds, and correlation model represented as black squares. The clustering structure is not that clear in terms of $b^0$ but the dissimilarities between networks constructed from different models are in general much higher than that between networks from the same model, and an unsupervised classification with one linear boundary in the embedded space would yield 10 out of 150 errors ($6.7\%$).

Next we consider networks with number of nodes ranging between $30$ and $39$. Ten networks are randomly generated for each network type and each number of nodes, resulting in $150$ networks in total. Figure \ref{fig_synthetic} (c) and (d) illustrate the two dimensional Euclidean embeddings of the network metric lower bounds $b^0$ and $b^1$ for these networks. Despite the fact that networks with same model have different number of nodes, dissimilarities between persistent homologies are smaller when their underlying networks are from the same process. Besides, networks with the correlation model are highly similar to each other regarding their corresponding persistent homologies, irrespective of the sizes of the networks nor the dimension of persistence diagrams. An unsupervised classification with one linear boundary would yield 6 out of 150 errors ($4.0\%$) for $b^0$ and 7 errors ($4.7\%$) for $b^1$. The performance of $b^0$ stays relatively unchanged when the number of nodes in the networks considered reside in a range of nodes. The simulations illustrate the applicability of persistent homology in identifying the patterns of networks of different processes. Similar results are obtained when different parameters are used in generating the networks. 

Here we give interpretations of why persistent homologies succeed in network discrimination. Since only pairwise relationships are examined, the bottleneck distance $b^0$ is determined by the death time of the $0$-dimensional persistent homologies. If we consider that nodes connected by an edge of weight $w$ become members of the same community at time $w$, the death time in $\PL 0$ can be interpreted as the time instant when isolated nodes join the main community of the network. We focus our attention on the 150 networks with size ranging between $30$ and $39$. Each row in Figure \ref{fig_interpretation} (a) represents a network, and plots the death time in the $0$-dimensional persistence diagram $\PL 0$ of the filtration induced from that network in ascending order from left to right. The top 50 networks represent those with Erd\H os-R\' enyi model, the middle 50 are with the unit circle model, and the bottom 50 are with the correlation model. Each row in Figure \ref{fig_interpretation} (b) plots the death time in $\PL 0$, in descending order from left to right, of the induced filtration. It is not a direct mirror of Figure \ref{fig_interpretation} (a) because the size of the networks ranges from $30$ to $39$ and so the number of points in the persistence diagrams are different. The time instants when isolated nodes join the main community in the networks with the correlation model are concentrated in the interval $[0.25, 0.4]$. This is due to the fact that the linear correlation coefficient between two randomly generated feature vectors cannot be too positive nor too negative. Also, networks with unit circle model are different from those constructed with the Erd\H os-R\' enyi model because in the latter case, the distribution of death time of points in $\PL 0$ has heavy tail towards $1$, which results from the fact that some points in the unit circle may be far away from the main component and it requires larger distance for them to join the main component.

For higher order persistent homologies, $b^1$ is only affected by the birth time of the $1$-dimensional persistent homologies, which can be interpreted as the maximum of the three pairwise connections between three nodes and therefore the time instant when a `closely-connected' community is formed by the three nodes. Each row in Figure \ref{fig_interpretation} (c) and (d) exhibits the birth time in $\PL 1$, in descending and ascending order from left to right respectively, of the filtration induced from the corresponding network. The time instants when three nodes in the networks with correlation model form a closely connected triplet are highly focused in the interval of $[0.4, 0.6]$. This is due to the fact that the when the correlation coefficient $\rho_{ij}$ between features $\bbu_i$ and $\bbu_j$ are small and $\rho_{ik}$ between $\bbu_i$ and $\bbu_k$ is small, $\rho_{jk}$ between $\bbu_j$ and $\bbu_k$ cannot be too small. For networks with unit circle model, the time points for many nodes in the network forming a closely connected triplet are either smaller or larger compared to the Erd\H os-R\' enyi model. This is because if three points in the unit circle are close to each other, all the pairwise distances between them would not be high; otherwise the minimum inscribed circle of the three points would possess large radius, resulting in a high value on the maximum of the pairwise distances. Admittedly, other methods to compare networks may also succeed in distinguishing networks, after some proper treatment towards the issue of different sizes. Nonetheless, persistent homology method would be more universal, not only for the reason that it establishes a lower bound to the actual network metrics, but also since it provides a systematic way to analyze the formation of communities in a given network.

%
\begin{figure*}[t] 
\begin{minipage}[h]{0.33\textwidth}
    	\centering
	\includegraphics[trim=1.5cm 0.5cm 1.5cm 0.5cm, clip=true, width=1 \textwidth]{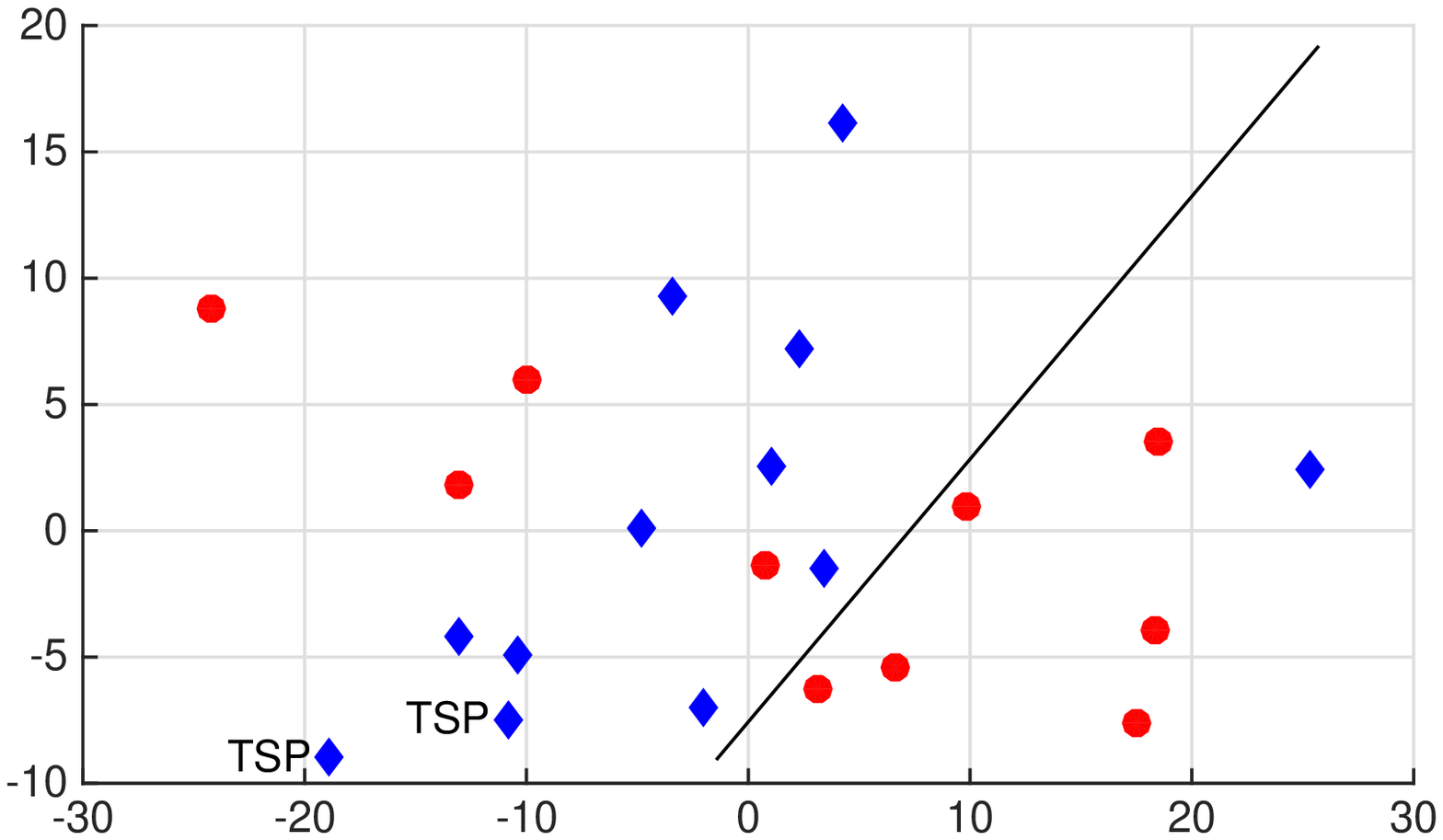}
	\scriptsize (b) $b^k(\PL 0)$
\end{minipage}
\begin{minipage}[h]{0.33\textwidth}
    	\centering
	\includegraphics[trim=1.5cm 0.5cm 1.5cm 0.5cm, clip=true, width=1 \textwidth]{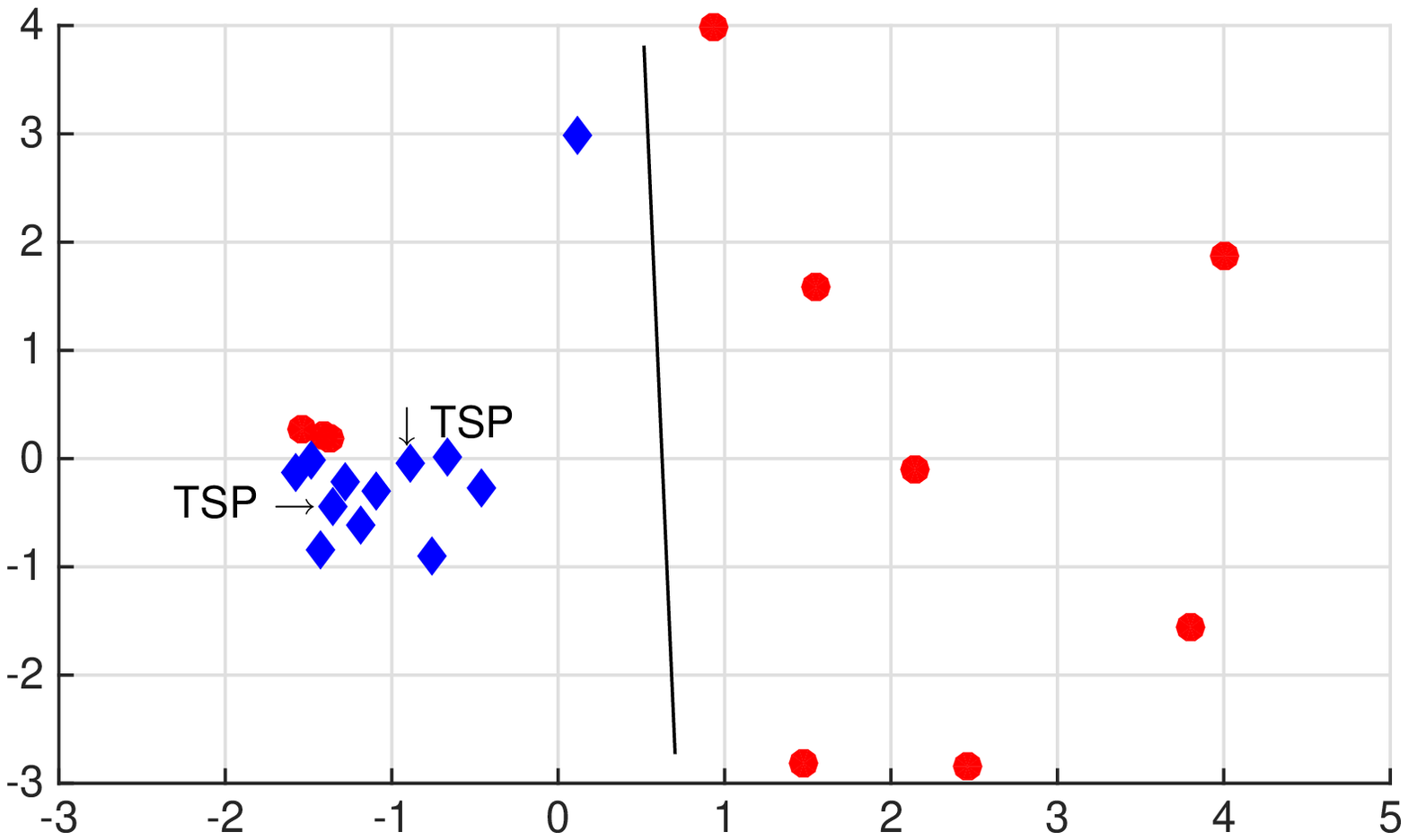}
	\scriptsize (c) $b^k(\PL 1)$
\end{minipage}
\begin{minipage}[h]{0.33\textwidth}
    	\centering
	\includegraphics[trim=1.5cm 0.5cm 1.5cm 0.5cm, clip=true, width=1 \textwidth]{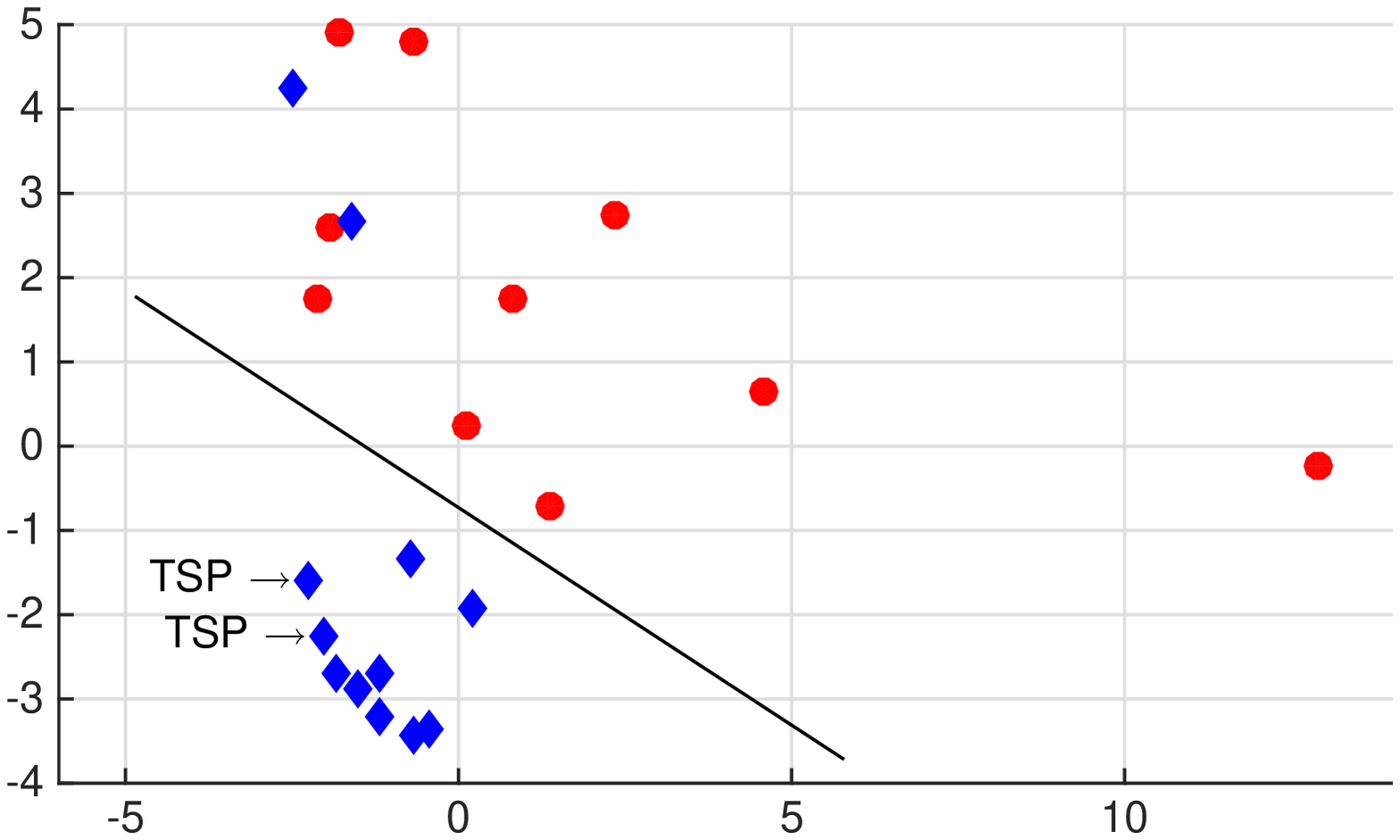}
	\scriptsize (d) $b^k(\PL 2)$
\end{minipage}
\caption{Two dimensional Euclidean embeddings of the networks constructed from quinquennial publications in engineering and mathematics journals with respect to the network metric lower bounds $b^k(\PL 0)$, $b^k(\PL 1)$, and $b^k(\PL 2)$. In the embeddings, red circles denote networks constructed from mathematics journals and blue diamonds represent networks from engineering journals. Networks constructed from publications of TSP are labeled.
\vspace{-3mm}
}
\label{fig_authorship}
\end{figure*}
%
%
\begin{figure*}[t] 
\begin{minipage}[h]{0.33\textwidth}
    	\centering
    	\includegraphics[trim=1.5cm 0.5cm 1.5cm 0.5cm, clip=true, width=1 \textwidth]{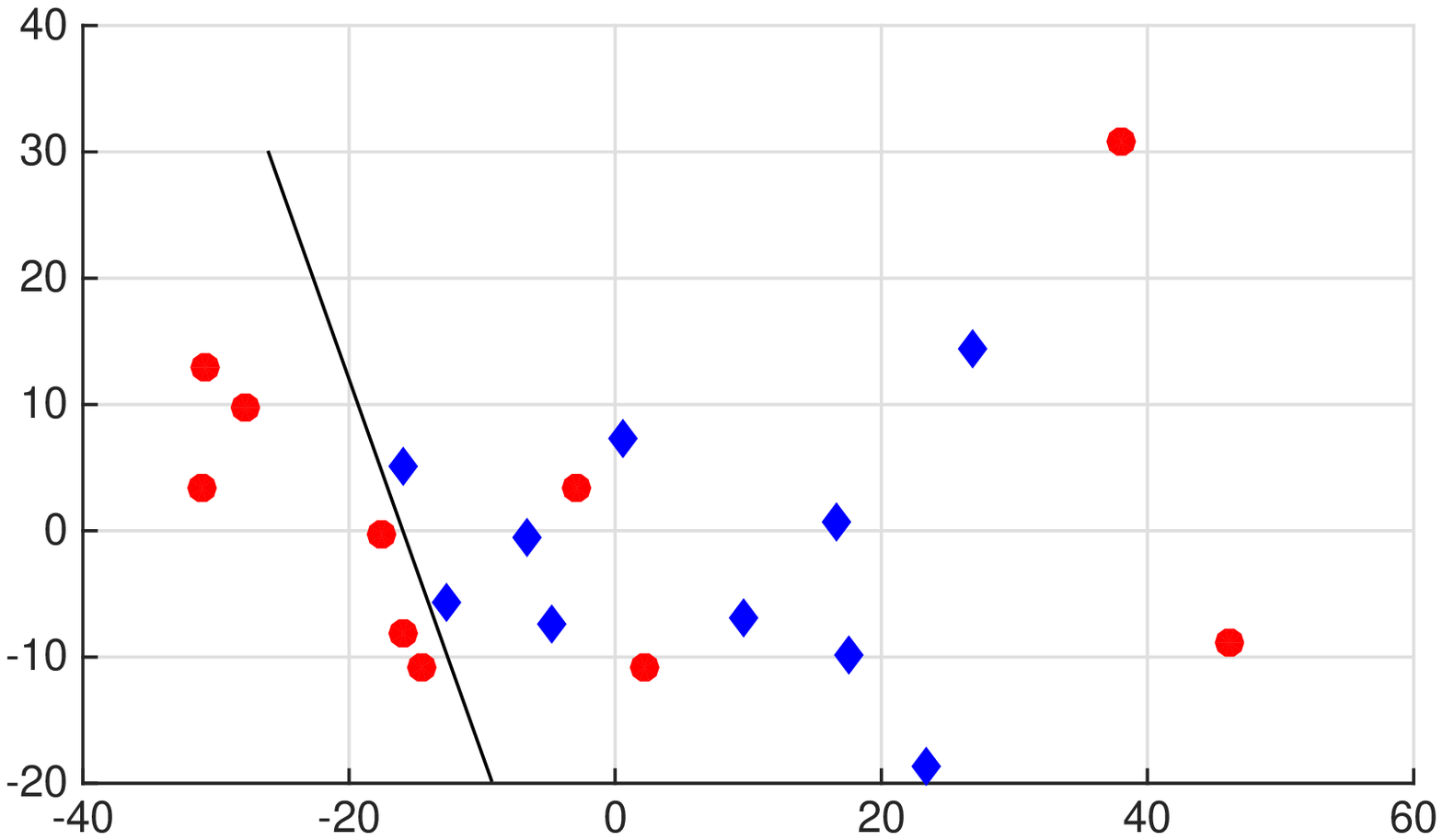}
	\scriptsize (a) TAC - TSP
\end{minipage}
\begin{minipage}[h]{0.33\textwidth}
    	\centering
	\includegraphics[trim=1.5cm 0.5cm 1.5cm 0.5cm, clip=true, width=1 \textwidth]{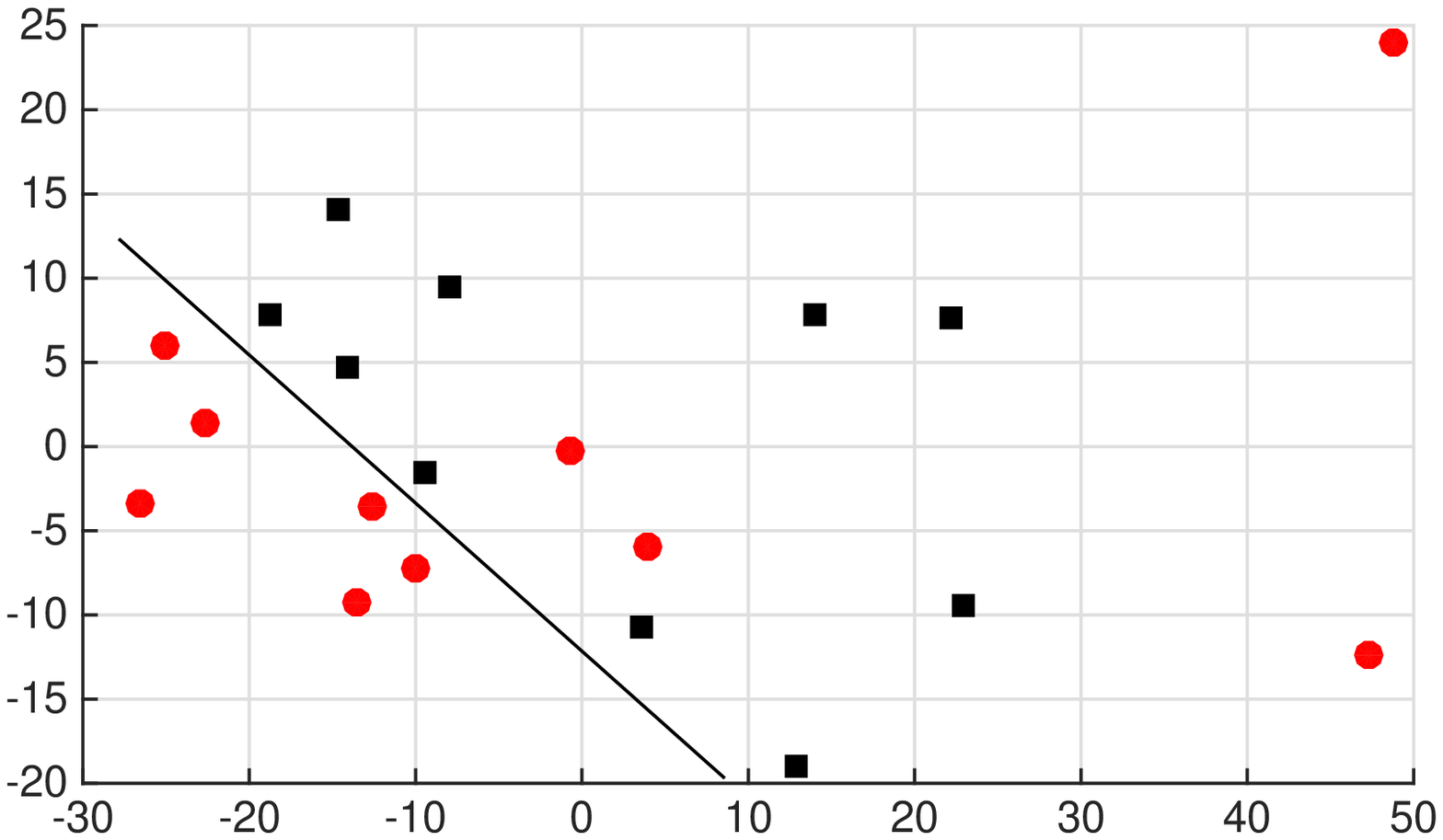}
	\scriptsize (b) TAC - TWC
\end{minipage}
\begin{minipage}[h]{0.33\textwidth}
    	\centering
	\includegraphics[trim=1.5cm 0.5cm 1.5cm 0.5cm, clip=true, width=1 \textwidth]{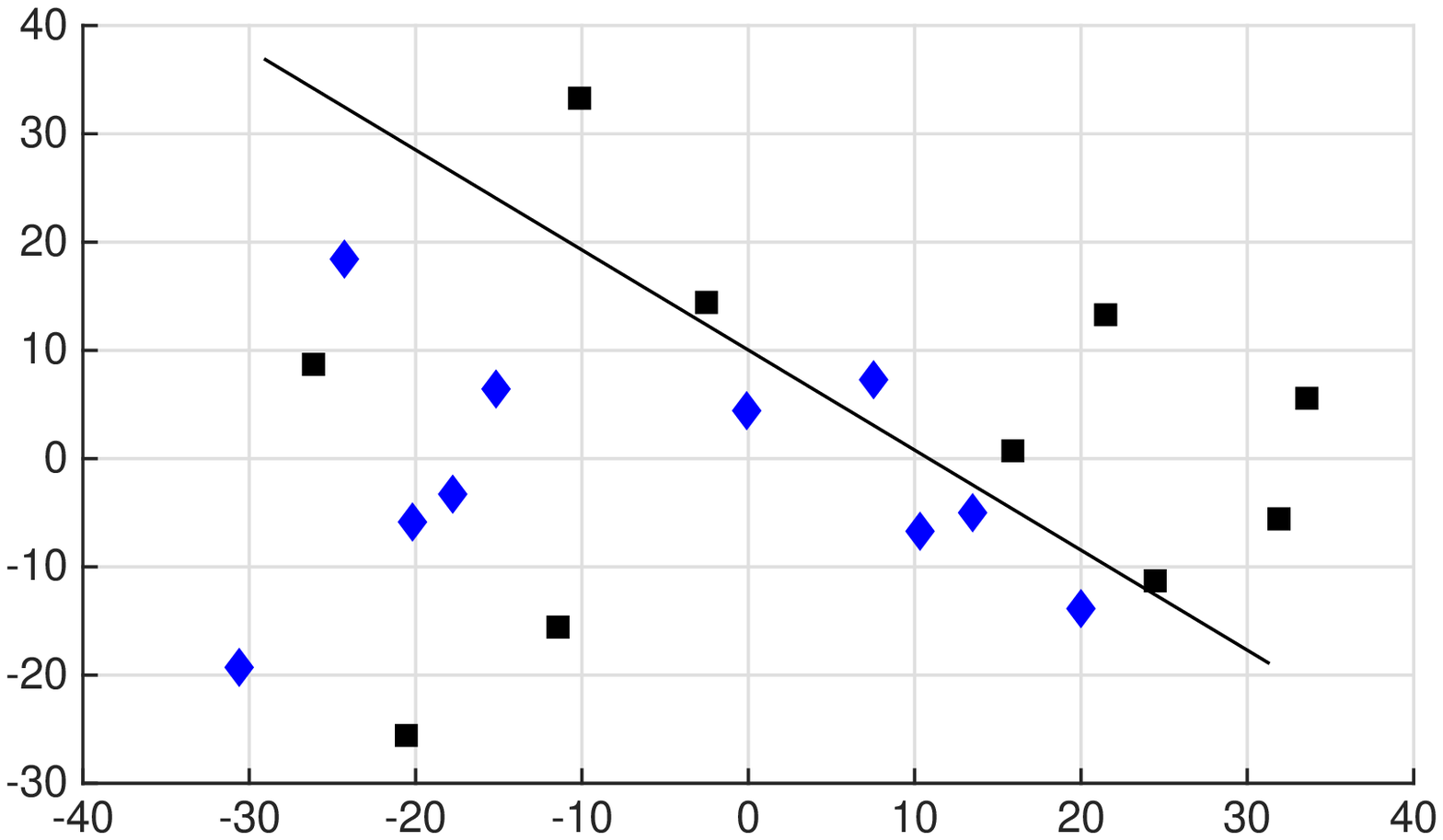}
	\scriptsize (c) TSP - TWC
\end{minipage}
\caption{Two dimensional Euclidean embeddings of the networks constructed from annual publications in TAC, TSP, and TWC with respect to the summation of the metric lower bounds $b^k(\PL 0)$, $b^k(\PL 1)$, and $b^k(\PL 2)$. In the embeddings, red circles represent TAC, blue diamonds TSP, and black squares TWC.\vspace{-4mm}
}
\label{fig_simulation_ASW}
\end{figure*}

%
\subsection{Comparison of Coauthorship Networks}\label{sec_compare_coauthorship}
We apply the lower bounds to compare $2$-order coauthorship networks where relationship functions denote the number of publications of single authors, pairs of authors, and triplets. These coauthorship networks are proximity networks because they satisfy the order decreasing property. We consider publications in $5$ journals from mathematics community: Computational Geometry (CG), Discrete Computational Geometry (DCG), J. of Applied Probability, (JAP) J. of Mathematical Analysis and Applications (JMAA), SIAM J. on Numerical Analysis (SJNA), and $6$ journals from engineering community, all from IEEE: Signal Processing Magazine (SPM), Trans. Automatic Control (TAC), Trans. Pattern Analysis and Machine Intelligence (TPAMI), Trans. Information Theory, Trans. Signal Processing (TSP), Trans. Wireless Communication (TWC). For each journal, we construct networks for the 2004-2008 and 2009-2013 quinquennia. For TAC, TSP, and TWC, we also construct networks for each annual from 2004 to 2013. Lists of publications are queried from \cite{EngineeringVillage}.

For each of these journals we consider all publications in the period of interest and construct proximity networks where the node set $X$ is formed by all authors of the publications. Zeroth order proximities are defined as the total number of publications of each member of the network, first order proximities as the number of papers coauthored by pairs, and second order proximities as the number of papers coauthored by triplets. To make networks with different numbers of papers comparable we normalize all relationships by the total number of papers in the network. The positive constant $\epsilon$ as in Figure \ref{fig_proximity_network_example} is set to $1/1000$. There are papers with more than three coauthors but we don't record proximities of order higher than 2. By assuming that networks from the same community or constructed from the same journal have similar collaboration patterns, we show here that network metric lower bounds succeed in identifying these patterns and in distinguishing coauthorship networks from communities with different interests.

Figure \ref{fig_authorship} shows the two dimensional Euclidean embeddings of the network metric lower bounds $b^0$, $b^1$ and $b^2$. The 12 engineering networks (blue diamonds) separate clearly from the 10 mathematics networks (red circles) in $b^1$ and $b^2$. The clustering is not that clear in $b^0$ but still networks from same community tend to be similar to each other. An unsupervised classification with one linear boundary running across the embeddings would generate errors of $2$ ($9.09\%$) to $5$ ($22.73\%$) out of $22$ networks. Networks constructed from the same journal tend to be close in the lower bounds. As an example, the networks of TSP with different quinquennia are marked in the embeddings and it is clear that their differences in homologies are considerably low. Such scenarios are observed for several other journals as well. 

We analyze the persistent homology of each of the coauthorship networks to investigate the reason why persistent homologies succeed in network discrimination. Compared to networks from mathematics communities, networks from engineering communities in general would yield $0$-dimensional persistent homologies with smaller birth time but larger death time, $1$-dimensional homologies with larger birth and death time, and $2$-dimensional homologies with larger birth time. An interpretation of such observations would be that in engineering, there exist more small communities that never collaborate with each other and it is uncommon to have a ``club'' of 3 to 5 authors in engineering that a strong collaboration exists between any pairs of the authors in the ``club''; such scenarios are absent for mathematicians.

As a comparison, we applied some simple and reasonable methods to compare the coauthorship networks considered in this section. Motifs have been shown effective in distinguishing coauthorship networks from different scientific fields \cite{Choobdar12}. To compare high order coauthorship networks by motifs, we restrict attention to pairwise relationships. The dissimilarities between coauthorship networks are assigned as the differences between the summations of the weighted motifs in their corresponding pairwise networks. Analysis based on triangle motifs (weighted) results in a clear cluster between networks from CG, DCG, JAP and another cluster between networks from TSP, TWC, but cannot distinguish other networks very well. Tetrahedron motif analysis (weighted) results in three clear clusters: networks from CG, DCG, JAP, networks from JMAA, SJNA, SPM, TPAMI, and networks from TSP, TWC. Other simple and common methods to compare pairwise networks yield similar results. Methods to compare networks via features give us similar observations as those based on the persistent methods; feature methods would generate 6 to 8 errors in classifications.

%
%
\subsection{Engineering communities with different research interests} \label{sec_engineering_different}
The network metric lower bounds succeed in distinguishing the different collaboration patterns in engineering and mathematics communities. We now illustrate that the lower bounds are also able to identify distinctive features of engineering communities with different research interests. To see this we consider the networks constructed from annual publications of TAC, TSP, and TWC. 

Figure \ref{fig_simulation_ASW} shows the two dimensional Euclidean embeddings of the networks with respect to the summation of the lower bounds $b^0$, $b^1$, and $b^2$. We expect more variations in annual networks because the time for averaging behavior is reduced. Besides, it is hard to argue that intrinsic and obvious differences exist in the collaboration patterns in automatic control, signal processing, and wireless communication communities. Still, networks constructed from the same journal but different annuals tend to be close to each other and form clustering structures. An unsupervised classification with one linear boundary in the embeddings run across the summation of lower bounds would generate $4$ ($20\%$) errors out of $20$ networks in all three classification problems considered. The less obvious clustering structure formed by networks from different journals in Figure \ref{fig_simulation_ASW} (c) compared to (a) and (b) also suggests that the collaboration patterns in research communities of signal processing and wireless communication are more similar compared to that of automatic control.

%
\section{Conclusion}\label{sec_conclusion}
We establish connections between high order networks and simplicial complexes and use the differences between the induced homological features to evaluate the differences between networks. We justify that this is a lower bound to two families of valid metrics in the space of high order networks modulo permutation isomorphisms. These lower bounds succeed in classifying weighted pairwise networks constructed from different processes, in distinguishing the collaboration patterns of engineering communities from mathematics communities, and in discriminating engineering communities with different research interests.



\appendices

%
\section{Proofs of Proposition \ref{prop_persistence_all}}\label{apx_proof_info}

Given any tuple $x_{0:k}$ with non-duplicating nodes, \eqref{eqn_proof_filtration_final} indicates that the $k$-simplex $\phi^k$ defined by the convex hull $\conv\{x_{0:k}\}$ appears strictly after any of its faces $\conv\{x_{0:\widehat s:k}\}$ in the filtration. Suppose $\phi^k$ appears at time $\alpha$ and denote $\partial_k \phi^k = \sum_i \beta_i \psi_i^{k-1}$ with $\beta_i$ the coefficients, then each $\psi_i^{k-1}$ appears strictly before time $\alpha$.

Now suppose that the appearance of $\phi^k$ trivializes a $(k-1)$-th dimensional homological feature. This means that $\phi^k$ is the boundary to trivialize the $(k-1)$-th dimensional cycle $\partial_k\phi^k$. Since each face $\psi_i^{k-1}$ of $\phi^k$ appears strictly before time $\alpha$, the cycle $\partial_k\phi^k$ results in a homological feature. The death time of this homological feature is $\alpha$, or equivalently, the time represented by the relationship $r_X^k(x_{0:k})$.

On the other hand, if the appearance of $\phi^k$ does not trivialize a $(k-1)$-th dimensional homological feature, then the $(k - 1)$-cycle $\partial_k \phi^k$ is in the collection of simplices appearing before or on time $\alpha$. This means that $\partial_k \phi^k$ can be represented by a sum of the boundaries of some $k$-chains $\Phi_i^k$,
\begin{align}\label{eqn_proof_all_other_cycles}
    \partial_k \phi^k = \sum_i \beta_i \partial_k\Phi_i^k
\end{align}
with coefficients $\beta_i$ and $k$-chains $\Phi_i^k$ appearing before or on time $\alpha$. By the definition of $k$-chains, $\Phi_i = \sum_j \beta'_j \psi_j^k$ with coefficients $\beta'_j$ and $k$-simplices $\psi_j^k$ appears before or on time $\alpha$. Therefore, \eqref{eqn_proof_all_other_cycles} can be written as $\partial_k \phi^k = \sum_j \beta''_j \partial_k\psi_j^k$. Rearranging terms,
\begin{align}\label{eqn_proof_all_identify_cycle}
         \partial_k \Big(\sum_j \beta''_j \psi_j^k - \phi^k \Big) = 0.
\end{align}
This implies that $\sum_i \beta''_i \psi_i^k - \phi^k$ is a $k$-cycle. There must be a new cycle formed since $\phi^k$ just appears. The cycle cannot be trivialized immediately since any $(k+1)$-chain $\Psi^{k+1}$ with $\partial_{k+1} \Psi^{k+1} = \sum_i \beta''_i \psi_i^k - \phi^k$ would involve a simplex $[x_{0:k, l}]$ for some node $x_l$ with tuple $x_{0:k, l}$ consisted of non-repeating elements where this simplex $[x_{0:k, l}]$ appears strictly after $\alpha$. Therefore we have a $k$-th dimensional homological feature with birth time $\alpha$, or equivalently, the time denoted by the relationship $r_X^k(x_{0:k})$. This concludes the proof.

%
\section{Proofs of Fact \ref{prop_augmented_network} and Fact \ref{prop_d_relationship}}\label{apx_proof_lower}

%
\begin{myproof}[of Fact \ref{prop_augmented_network}]
We need to prove the (i) symmetry property and (ii) identity property of high order networks, (iii) $\ccalL(A_X^K)$ being a valid filtration, and (iv) $\ccalL(A_X^K)$ and $\ccalL(D_X^K)$ having identical persistent homology.

%
\begin{myproof}[of the symmetry property]
For any tuples $a_{0:k}$, the symmetry property of dissimilarity network $D_X^K$ and the definition of augmented networks imply $r_{A_X}^k(a_{[0:k]}) = r_X^k(x_{[C_0:C_k]}) = r_X^k(x_{C_0:C_k}) = r_{A_X}^k(a_{0:k})$ for any reordering $a_{[0:k]}$. This shows the symmetry property of $A_X^K$.
\end{myproof}

%
\begin{myproof}[of the identity property]
Given a tuple $a_{0:k}$, if its subtuple $a_{l_0: l_\tdk}$ have same set of unique elements as that of $a_{0:k}$, according to the Definition of augmented networks, we would have that $x_{C_0:C_k}$ and $x_{C_{l_0}:C_{l_\tdk}}$ also possess identical set of unique elements. The identity property of dissimilarity network $D_X^K$ and the definition of augmented networks yield $r_{A_X}^k(a_{0:k}) = r_X^k(x_{C_0:C_k}) = r_X^\tdk(x_{C_{l_0}:C_{l_\tdk}}) = r_{A_X}^\tdk(a_{l_0:l_\tdk})$. This shows the identity property of $A_X^K$.
\end{myproof}

%
\begin{myproof}[of $\ccalL(A_X^K)$ being a valid filtration]
The order increasing property of relationship function in augmented networks holds true due to the fact that
\begin{align}\label{eqn_proof_augmented_filtration}
r_{A}^k(a_{0:k})\!=\! r_X^k(x_{C_0:C_k}) \!\geq\! r_X^{k-1}(x_{C_0:C_{k-1}}) \!=\! r_{A}^{k-1}(a_{0:k-1}).
\end{align}
The remaining proof is identical to the proof of Proposition \ref{prop_filtration}. \end{myproof}

%
\begin{myproof}[of $\ccalL(A_X^K)$ and $\ccalL(D_X^K)$ having identical persistence intervals]
First, for each point $x \in X$, pick one pair $(x, y)$ from the correspondence $C$ to construct $C_0$ that is a subset of $C$. If we define a map that maps each pair $(x, y)$ to its first element $x$, this gives a bijective projection from $C_0$ to $X$. Construct $A_{C_0, X}^K$ and $A_{C, X}^K$ as the augmented networks using the respective correspondence. It then follows naturally that $A_{C_0, X}^K$ is isomorphic to the the original dissimilarity network $D_X^K$ and so do the corresponding filtrations. Denote $\ccalL := \ccalL(A_{C, X}^K)$ and $\ccalL_0 := \ccalL(A_{C_0, X}^K) = \ccalL(D_X^K)$.

Next, consider the projection of the filtration $\ccalL$ onto $\ccalL_0$, where each vertex $(x, y)$ in $C$ is mapped to $(x, y_0)$ in $C_0$ who share the first element in the pair. Denote $L^\alpha$ and $L^\alpha_0$ as the simplicial complexes that collect simplices appearing before or prior to $\alpha$ in the respective filtration. For each value of $\alpha$, the projection defines a retraction of simplicial complexes $L^\alpha \rightarrow L^\alpha_0$. Since the relationship functions are the same in the augmented network $A_{C, X}^K$ as in the original network $A_{C_0, X}^K$, whenever a simplex $\sigma$ appears in $L^\alpha$, not only its projection appears in $L_0^\alpha$ at the same time, but all the simplices connecting vertices of $\sigma$ and vertices of its projection are also already in $L_0^\alpha$. Hence, as a simplicial map, the projection $L^\alpha \rightarrow L_0^\alpha$ is contiguous to the identity of $L^\alpha$. As a result, the section $L_0^\alpha \rightarrow L^\alpha$ is a homotopy equivalence. Its induced homomorphism at the homology level is therefore an isomorphism, and consequently the two persistence diagrams are isomorphic.
\end{myproof}

Having demonstrated all statements, the proof completes. \end{myproof}

%
\begin{myproof}[of Fact \ref{prop_d_relationship}]
We need to prove the two statements in \eqref{eqn_prop_d_relationship} and \eqref{eqn_prop_d_relationship_2}. Given any $0 \leq k'\leq k\leq K$, it follows from \eqref{eqn_d_N_prelim} that for any correspondence $C$ between $X$ and $Y$,
\begin{align} 
\label{eqn_proof_d_rela_start_k}
   \Gamma_{X,Y}^k (C) &= \max_{(x_{0:k}, y_{0:k}) \in C}
            \left| r_X^k(x_{0:k}) - r_Y^k (y_{0:k}) \right|, \\
\label{eqn_proof_d_rela_start_k_prime}
   \Gamma_{X,Y}^{k'} (C) &= \max_{(x_{0:k'}, y_{0:k'}) \in C}
            \left| r_X^{k'} (x_{0:k'}) - r_Y^{k'} (y_{0:k'}) \right|.
\end{align}
For the $(x^\star_{0:k'}, y^\star_{0:k'}) \in C$ achieving the maximum difference $| r_X^{k'} (x^\star_{0:k'}) - r_Y^{k'} (y^\star_{0:k'}) |$ in $\Gamma_{X,Y}^{k'} (C)$, we can construct another correspondent pair $(x^\star_{0:k}, y^\star_{0:k})$ such that $x^\star_{0:k'}$ is a sub-tuple of $x^\star_{0:k}$ with identical set of unique elements and $y^\star_{0:k'}$ is a sub-tuple of $y^\star_{0:k}$ with identical set of unique element. It follows from the identity property of high order networks that $r_X^{k'}(x^\star_{0:k'}) = r_X^k(x^\star_{0:k})$ and $r_Y^{k'}(y^\star_{0:k'}) = r_Y^k(y^\star_{0:k})$. This implies that taking the maximum $|r_X^k (x_{0:k}) - r_Y^k (y_{0:k})|$ over ${(x_{0:k}, y_{0:k}) \in C}$ cannot yield a lower difference than $|r_X^{k'} (x^\star_{0:k'}) - r_Y^{k'} (y^\star_{0:k'})|$, i.e.
\begin{align}\label{eqn_proof_d_rela_Gamma}
    \Gamma_{X,Y}^k (C) \geq \left| r_X^{k'} (x^\star_{0:k'}) - r_Y^{k'} (y^\star_{0:k'}) \right|
        = \Gamma_{X,Y}^{k'} (C).
\end{align}
Since \eqref{eqn_proof_d_rela_Gamma} holds true for any correspondence $C \in \ccalC(X, Y)$, the inequality must hold true when we take the minimum over all correspondences $\ccalC(X, Y)$ between $X$ and $Y$,
\begin{align}\label{eqn_proof_d_rela_Gamma_min}
    \min_{C \in \ccalC(X, Y)} \Gamma_{X,Y}^k (C) \geq \min_{C' \in \ccalC(X, Y)}\Gamma_{X,Y}^{k'} (C').
\end{align}
Substituting the definition of $k$-order and $k'$-order network distances into \eqref{eqn_proof_d_rela_Gamma_min} yields $d_\ccalD^k \geq d_\ccalD^{k'}$, concluding the proof of \eqref{eqn_prop_d_relationship}.

Also, it follows from \eqref{eqn_d_N_norm_prelim} that for any correspondence $C$ between the node sets $X$ and $Y$, the network difference between $N_X^K$ and $N_Y^K$ measured by $C$ is 
\begin{align}\label{eqn_proof_d_rela_infty_start}
    \left\| \bold \Gamma_{X, Y}^K(C) \right\|_\infty = \ \max_{k = 0, 1, \dots, K} \left\{ \Gamma_{X, Y}^k(C) \right\}.
\end{align}
From \eqref{eqn_proof_d_rela_Gamma} we know that for any $k'$, $\Gamma_{X,Y}^k (C) \geq \Gamma_{X,Y}^{k'} (C)$, and therefore for any correspondence $C$, $\max_{k = 0}^K \left\{ \Gamma_{X, Y}^k(C) \right\} = \Gamma_{X, Y}^K(C)$. Substituting this into \eqref{eqn_proof_d_rela_infty_start} and taking a minimum over all correspondences concludes the proof of \eqref{eqn_prop_d_relationship_2}. 

Having proven the two statements, the proof completes.\end{myproof}

%
\urlstyle{same}
\bibliographystyle{IEEEtran}
\bibliography{barcodes_biblio}

\end{document}